\documentclass[structabstract]{aa}  
\usepackage[T1]{fontenc}
\usepackage{amsmath}
\usepackage{natbib}
\usepackage{graphicx}
\usepackage{txfonts}

\begin{document}
   \title{Signatures of massive collisions in debris discs}
   \subtitle{A self-consistent numerical model}
   \author{Q. Kral
          \inst{1},
          P. Th\'ebault\inst{1},
          J.-C. Augereau\inst{2,3},
          A. Boccaletti\inst{1},
          S. Charnoz\inst{4}
          }
   \institute{LESIA-Observatoire de Paris, UPMC Univ. Paris 06, Univ. Paris-Diderot, France
    	\and
	Univ. Grenoble Alpes, IPAG, F-38000 Grenoble, France 
	\and
	CNRS, IPAG, F-38000 Grenoble, France 
	\and
	    Laboratoire AIM, Universit\'e Paris Diderot / CEA / CNRS, Institut Universitaire de France
             }
  
\offprints{Q. Kral} \mail{quentin.kral@obspm.fr}
\date{Received ; accepted } \titlerunning{Signatures of massive collisions in debris discs}
\authorrunning{Kral}

 
  \abstract
   {Violent stochastic collisional events have been invoked as a possible explanation for some debris discs displaying pronounced azimuthal asymmetries or having a luminosity excess exceeding that expected for systems at collisional steady-state. So far, no thorough modelling of the consequences of such stochastic events has been carried out, mainly because of the extreme numerical challenge of coupling the dynamical and collisional evolution of the released dust.}
   {We perform the first fully self-consistent modelling of  the aftermath of massive breakups in debris discs. We follow the collisional and dynamical evolution of dust released after the breakup of a Ceres-sized body at 6 AU from its central star. We investigate the duration, magnitude and spatial structure of the  signature left by such a violent event, as well as its observational detectability. }
   {We use the recently developed LIDT-DD code (Kral et al., 2013), which handles the coupled collisional and dynamical evolution of debris discs. The main focus is placed on the complex interplay between destructive collisions, Keplerian dynamics and radiation pressure forces. We use the GRaTer package to estimate the system's luminosity at different wavelengths.}
   {The breakup of a Ceres-sized body at 6\,AU creates an asymmetric dust disc that is homogenized, by the coupled action of collisions and dynamics, on a timescale of a few $10^5$ years. The particle size distribution in the system, after a transient period where it is very steep, relaxes to a collisional steady-state law after $\sim 10^4$ years. The luminosity excess in the breakup's aftermath should be detectable by mid-IR photometry, from a 30\,pc distance, over a period of $\sim 10^6$ years that exceeds the duration of the asymmetric phase of the disc (a few $10^5$ years). As for the asymmetric structures, we derive synthetic images for the SPHERE/VLT and MIRI/JWST instruments, showing that they should be clearly visible and resolved from a 10\,pc distance. Images at 1.6$\mu$m (marginally), 11.4 and 15.5$\mu$m would show the inner disc structures while 23$\mu$m images would display the outer disc asymmetries.}
   {}

   \maketitle
%

\section{Introduction}

Understanding the origin of the spatial structures observed in most resolved debris discs is paramount for our comprehension of planetary system evolution and formation, as these structures could be the signature of planet perturbations, stellar companions, ISM interactions, or interaction with gas \citep[see for example, the reviews by][]{wyat08, kriv10}. 

Another possible disc-sculpting mechanism is the breakup of large planetesimals following violent collisions. Such violent events are indeed expected during the late stages of planet formation ($\sim$ 10-100 Myrs), which should be a highly chaotic period where large planetary embryos interact through high-velocity collisions. This is especially true for the terrestrial planet regions, where large scale collisions should be the most frequent \citep{keny05,keny06,raym09}. Even at much later periods, such violent stochastic events should still occur, as evidenced by the existence of asteroid families, which are the result of the breakup of large parent bodies and whose age is much smaller than that of the solar system \citep[e.g.][]{zapp02,durd07}.

Such violent collisional events have been invoked as a possible explanation for some debris discs exhibiting unusual characteristics. It has for instance been considered for discs that appear much too luminous for their luminosity to be explained by a steady-state collisional cascade \citep{wyat08,gasp13}. For some systems, massive breakups have also been invoked to explain the presence of large amounts of warm dust that cannot be sustained at this level for the system's age. A good example of these bright warm dust discs is the $\sim$ 12 Myrs old \object{HD172555} system, for which the occurrence of a recent massive collision is furthermore supported by the detection of glassy silica and possibly SiO vapors, which could have formed at high temperatures in the aftermath of a giant hypervelocity impact \citep{liss09,john12}. Similarly, massive collisions could also explain the large amounts of hot short-lived sub-micron grains detected in some discs such as the $\sim$ 10 Myrs old \object{HD113766} system \citep{olof13}.
Massive breakups have also been considered as a possible source for some bright features observed in resolved discs. \citet{wyat02} have examined if collisionally-produced dust clumps could be observed around Fomalhaut, while \citet{tele05} proposed that some mid-IR asymmetries observed in the $\sim$ 12 Myrs old \object{$\beta$ Pictoris} disc could be the result of a cataclysmic planetesimal breakup. This interpretation of the Telesco's et al. clump might be supported by a recent ALMA detection of a CO clump in the same region \citep{dent14}.

On the theoretical side, stochastic massive breakups in debris discs have been investigated in several numerical studies over the past decade, but so far only with codes relying on simplified modelling of the system's physics. The main challenge is here to simultaneously model the dynamical and collisional evolution of the system, firstly because collisions in debris discs are expected to be destructive and fragment-producing, but also because the dynamics of these small fragments is highly size-dependent due to stellar radiation pressure \citep[see discussion in][]{theb12b}. This is the reason why, historically, the dynamical and collisional evolution of debris discs had been studied separately, with deterministic $N$-body codes and statistical particle-in-a-box models, respectively. 

One of the first attempts at modelling large-scale breakups in debris discs is that of \citet{keny05}, who used their coagulation code \citep[e.g.][]{keny02,keny04} to follow the fate of collisions amongst the largest bodies of their accreting/fragmenting planetesimal swarm. They concluded that collisions amongst 100-1000km objects could be observable at mid-IR wavelengths as bright clumps or rings of dust, and used simple $N$-body simulations to estimate the survival time, $\sim\,$100 orbits, of such clumps due to Keplerian shear. However, these results were in essence 1-D (radial) and the coupling between dynamics and collisions was very partial. 
The more sophisticated study by \citet{jack12} investigated one specific massive breakup event, the Moon-forming impact, using an $N$-body code to follow the dynamical fate of the initially produced fragments and plugging in an analytical collisional mass removal rate. They concluded that such a Moon-forming impact would have been readily detectable around other stars in \emph{Spitzer} 24$\mu$m surveys for around 25 Myr. However, the collisional evolution of the system was followed assuming an axisymmetric disc and that a steady-state equilibrium collisional cascade holds at all time. 
Using a similar $N$-body based approach, \citet{jack14} investigated the observability of debris from giant impacts at large orbital radii and concluded that the resulting asymmetric discs might be observable for around 1000 orbital periods. However, collisions were here also treated in a simplified way, by estimating a system-integrated lifetime of the largest debris used to derive the global mass loss of the disc (here again assuming that the same steady-state collisional cascade holds everywhere and at all times). Furthermore, the coupled effect of collisions and radiation pressure, which is crucial for the evolution of the small grains dominating the flux at near-to-mid-IR wavelengths \citep{theb14}, was not taken into account.

We aim to reinvestigate these issues, using the newly developed LIDT-DD algorithm \citep[][ hereafter KTC13]{kral13}, which is the first debris disc code to take into account the coupled effect of dynamics and collisions in a fully self-consistent way. 

Achieving this coupling has been the goal of several numerical efforts over the past decade. After the pioneering works of \citet{grig07} and \citet{char03,char07}, the CGA algorithm of \citet{star09} and the DyCoSS code of \citet{theb12} have achieved a partial coupling of dynamics and collisions. These codes have given important results for the study of Kuiper Belt dust as well as discs in binaries, but they are restricted to systems at steady state and cannot follow second-generation collision fragments. The LIPAD \citep{levi12} and SMACK \citep{nesv13} codes do not suffer from these limitations. However, SMACK relies on the restrictive assumption that dynamics is not size-dependent, thus implicitly neglecting the crucial effect of radiation pressure. This simplification prevents it from modelling particles typically in the < 100$\mu$m size range. As for LIPAD, it includes size-dependence of the dynamics, but it has been designed to study the evolution of large planetesimals and assumes that all ``dust'' is regrouped into one single size-bin. Because of these restrictions, both codes cannot accurately model the grain size range that dominates debris disc luminosities at most observed wavelengths \citep[e.g.,][]{theb07}.

LIDT-DD's hybrid structure allows to overcome these restrictions and can handle the specificities of the dust physics in debris discs. It is in particular able to treat fragmenting collisions and take into account the crucial size dependence of the dynamics induced by radiation pressure.
The possibilities offered by LIDT-DD are well suited to the specific problem studied here, i.e., the highly collisional and spatially anisotropic event that is the breakup of a massive planetesimal. Our main objective is here to estimate the observability and the longevity of such an event. We investigate especially how the concurring effects of collisional activity and radiation pressure affect the asymmetric post-breakup structures. We focus on the inner regions of debris discs, around 6\,AU, because, according to our current understanding of the late stages of planet formation, massive breakups should be more frequent there \citep{keny04,keny05}.  We consider the case of a $\sim 500\,$km parent body, the typical size of the largest object in the present asteroid belt.

We present the main characteristics of LIDT-DD in section 2, highlighting the improvements that have been implemented for the present study, as well as the set-up considered for our nominal simulation. Section 3 presents the results obtained and investigates the longevity and detectability of the post-breakup debris cloud. We also derive synthetic images for the SPHERE/VLT and MIRI/JWST instruments at several near-to-mid infrared (IR) wavelengths. Section 4 explores the parameter dependence of the simulations and presents some simple scaling laws to extrapolate our results to alternative set-ups. Section 5 discusses the limitations of these LIDT-DD-based explorations. Conclusions and perspectives are given in the last Section.

\section{Model}\label{model}

For a full description of our code, we refer the reader to KTC13. Let us here briefly summarize its main characteristics as well as the main upgrades and improvements that have been implemented.

\subsection{Principle}

The basic principle of the LIDT-DD model is to couple a Lagrangian approach for the dynamics to a particle-in-a-box statistical Eulerian one for the collisional evolution \citep{char12}.
At a given location in the system, all particles of a given size $and$ sharing similar dynamical characteristics are gathered into larger super-particles (called ``tracers'')\footnote{Note that these super-particles are different from the ones used in the SMACK code, which each stands for a complete size-distribution because of the assumption that the dynamics is the same regardless of particle sizes.}, whose dynamical evolution is followed with an $N$-body scheme, while their collisional evolution is investigated with a statistical approach, considering all mutual tracer-tracer impacts at any given location in the system.

The procedure to evolve both the tracers' dynamics and collisions is then schematically the following:

\begin{itemize}
\item Each particle in the code is a super-particle representing a vast population of same-sized physical particles.
\item The positions and velocities of the tracers are integrated with a Bulirsh-St\"{o}er scheme ($N$-body approach) that is able to include different type of forces (Poynting-Robertson drag, radiation pressure, gravitational interactions, gas drag if needed). 
\item Once the dynamics has been integrated over one time step, the system is divided into spatial cells. The collisional evolution is then estimated cell by cell, by taking into account all potential tracer-tracer encounters within each cell. Although LIDT-DD is intrinsically 3-D and tracers dynamics is integrated in the vertical direction, we use a 2-D grid ($r$,$\theta$) for these ``collisional cells'', each individual cell having a finite vertical extension equal to the tracers' inclination times the radial distance.
\item All tracer-tracer collisions are then treated with a statistical procedure, taking into account the mutual velocities between tracers and the number density of real physical particles they represent. The size-distribution of the fragments produced by each of these impacts is estimated with the collisional outcome prescription used in the statistical code of \citet{theb07}. Its main parameter is the critical specific energy $Q^*$ required for dispersing at least 50\% of the target (see Table 1). Both fragmenting ($Q>Q^*$, where $Q$ is the collision kinetic energy per target unit mass), and cratering (or ``erosive'', $Q<Q^*$) impacts are taken into account.
The collisional debris are then redistributed, according to their sizes and dynamical characteristics, into newly-created tracers. The feedback of collisions onto the old and new tracers (momentum redistribution and energy loss) is taken into account.
\item After each dynamical+collisional time step, tracers of a given spatial cell are sorted into dynamical ``families'', in order not to lose important information about the dynamical complexity of the system (for instance, at a given location, grains having similar sizes can have different origins and thus different dynamical evolutions).
\item Finally, to avoid an unmanageable increase of the number of tracers, the code is looking, at the end of each time step, for redundant tracers which are then merged into the nearest tracers representing the same size and dynamical family.
\end{itemize}

\subsection{Upgrades}\label{upgrade}

Several updates have been implemented into LIDT-DD for the present study:

\begin{itemize}

\item \emph{Inclusion of a radiative transfer code}\\
The GRaTer radiative transfer code \citep{auge99,auge06,lebr13} has been fully coupled to LIDT-DD. It allows, from a given grain composition, stellar type and spatial location, to compute the absorption/emission and scattering coefficients for all the grains in the system. As a consequence, radiation pressure forces can be worked out self-consistently with GRaTer. Furthermore, GRaTer can be used to plot synthetic images at different wavelengths as well as the Spectral Energy Distribution (SED) of the system. \\

\item \emph{New dynamical sorting procedure for particles close to the blow-out limit}\\
Instead of the sorting procedure described in KTC13 (the clustering method of Ward), a more straightforward and accurate method is used for small grains strongly affected by radiation pressure. Indeed, such micron-sized grains have high eccentricities and special care must be taken to avoid merging intrinsically different dynamical families into a common tracer. The basic principle is the same as in KTC13 as we still sort the grains in a  $q+ a$ vs. $Q-a$ plane, where $a$ is the semi-major axis, $q$ and  $Q$ are the periastron and the apastron respectively. The difference is that, instead of identifying families using a clustering method, we define a 2-D sorting grid in the $(q+a,Q-a)$ plane and consider that a given dynamical family consists of all the tracers within a given cell of this grid. The grid can be either linear or logarithmic depending on the configuration. This procedure is applied to particles in the $0.05 < \beta < 0.5$ range, where $\beta$ is the size-dependent ratio between the radiation pressure and gravitational forces. \\

\item \emph{Adaptive time step}\\
In order to handle the strong time and location dependence of stochastic processes, an adaptive time step procedure has been implemented. This is a crucial upgrade for the earliest phases after the initial breakup, when variations of both the collision rates and the dynamical evolution can be very strong.\\

\item \emph{Fine size sampling close to the blow-out limit}\\
As already stated, debris disc luminosities are, in most cases, dominated by small bound grains close to the $\beta=0.5$ value. Great care has thus to be taken regarding the modelling of these grains, all the more because their dynamical evolution is extremely sensitive to small differences in sizes\footnote{The apastron of a particle created at a radial distance $r_0$ from parent bodies on a circular orbit is equal to $\frac{r_0}{1-2\beta}$, and has a very steep size-dependence close to $\beta=0.5$.}. 
As a consequence, a very fine sampling in size is used in the $0.05\leq \beta \leq 0.5$ domain, with a logarithmic size-increment  of $\epsilon_f=1.03$, instead of $\epsilon_c=1.6$ for tracers representing larger particles.
\end{itemize}

\subsection{Setup}\label{setup}

\begin{table*}
\caption{\label{tabtest}Relevant parameters used for the fiducial generic test run simulation in Sect.~\ref{results}}
\centering
\begin{tabular}{lc}
\hline\hline
Star\\
\hline
Spectral-type & A7V \\
Mass        & $1.84 \, M_\odot$\\
Magnitude in V    &  4.779 \\
Distance from observer & 30 pc \\
\hline
Grain physical characteristics\\
\hline
Blow-out size ($s_\mathrm{cut}$) & $1.8 \, \mu m$ \\
Material & Astrosilicate \\
Porosity & 0 \\
Density ($\rho$) & 2500 $\textrm{kg.m}^{-3}$ \\
\hline
Released fragment population\\
\hline
Minimum size     & $2.0 \, \mu$m ($\beta$ = 0.44) \\
Maximum size   & 1 m  \\
Initial size distribution & $\textrm{d}N \propto s^{-3.8} \textrm{d}s$ \\
Initial total mass & $10^{21}$ kg \\
Initial velocity dispersion & $0<v_\mathrm{frag}<v_\mathrm{esc}$\\
Initial release distance from the star & 6 AU \\
Initial mean eccentricity & 0.037 \\
\hline
LIDT-DD specific parameters\\
\hline
Number of initial tracers & 300 000 \\
Collisional grid cells number ($N_R$ x $N_\theta$) & 10 x 10 \\
Collisional grid cells spacing (log-scale) & $dr_\mathrm{int}=$1.5,  $dr_\mathrm{ext}=$13AU\\
\hline
Collisional prescription\\
\hline
Critical specific energy $Q^*$ & \citet{benz99} \\
$Q^*$ Formula & $Q^*=\alpha_1 \, \left(\frac{R_{{tar}}}{R_0}\right)^a+\alpha_2 \, \rho \,\left(\frac{R_{{tar}}}{R_0}\right)^b$\\
     & $a=-0.38$, $b=1.36$, $R_0$=1m \\
     & $\alpha_1=3.5 \times 10^{3}$J/kg, $\alpha_2=3 \times 10^{-8}$(SI)\\
Fragmentation & \citet{fuji77} \\
Cratering & \citet{theb07} \\
\hline
\end{tabular}
\end{table*}

We follow the evolution of a massive amount of small fragments, of mass $M_\text{frag} = 10^{21}$ kg, released by a violent phenomenon, in the inner regions (at $r_\text{init} = 6\,$AU) of a planetary system. In these inner regions, the most likely way of producing this massive release is the breakup of a massive body after a violent impact with a high-velocity projectile. We are not interested in investigating this violent event itself, but will choose initial parameters that are consistent with such a breakup. If we assume that $M_\text{frag}$ corresponds to the mass of the object that has been shattered, then this object's radius is $\sim 500\,$km, approximately the size of Ceres. We take this size as a reasonable order of magnitude for the typical size of the largest objects present in the late stages of planet formation.
\footnote{Note that \citet{jack14} considered a different kind of initial violent event, i.e. the cratering of a much larger planet-sized object, where only 3 per cent of the progenitor's mass escapes. This different choice was justified by the fact that these authors considered regions further out in the disc (50\,AU), where impacts are expected to be much less energetic and thus more likely to craterize than fully destroy the progenitor. And releasing large amounts of dust in such low energy impacts requires much larger progenitors than for fully destroying impacts closer to the star.}

Given that the breakup of such a large object requires an impact with a massive projectile, and that the probability for such two-body encounters is likely to decrease with object sizes \citep[because the size distribution of planetesimals is expected to be a function steeply decreasing with size, e.g.][]{keny04b}, we consider the least extreme case where the energy of the impact is close to the minimum one required to break up the target. This means that the velocities $v_\text{frag}$ of the escaping post-impact fragments, relative to the local Keplerian velocity, are not too high. We consider here that $v_\text{frag}$ are randomly distributed between 0 and $v_\text{esc}$, where $v_\text{esc}$ is the escape velocity of the initial target\footnote{Of course, the real velocity distribution of ejecta follows a more complex pattern \citep[e.g.,][]{lein12}, depending significantly on the characteristics of the initial breakup (relative sizes of the impactors, impact velocity), as well as varying with fragment sizes. However, it is not our goal to enter into the precise description of the initial breakup's outcome, and we take the $0\leq v_\text{frag}\leq v_\text{esc}$ criteria as a convenient way of parameterizing the impact's energy.}. Note that the fragments velocity is not the immediate post-breakup one, but is in principle the velocity at infinity, i.e., once fragments have escaped the initial body's gravity. We thus implicitly assume that all the considered fragments are gravitationally unbound to the initial target (in other words, their ``real'' immediate post-breakup velocity was between $v_\text{esc}$ and $2\,v_\text{esc}$). 
The initial velocity has then two components: 1) the Keplerian velocity of the progenitor parent body, and 2) a kick velocity, which has no reasons not to be isotropic \citep{jack12}, with values randomly chosen between 0 and $v_\text{esc}$, and kick angles isotropically distributed onto a sphere. 
The initial eccentricity and inclination distributions of the ejecta are then automatically obtained from this constraint on $v_\text{frag}$. For our nominal set-up, we obtain $<e>\, = 2 <i> \, \sim 0.037$.
The role of the initial fragment velocity will be explored by running additional simulations with higher values of $v_\text{frag}$ (Sect.~\ref{param}).

As an initial location for the dust release, we take 6\,AU. This choice is motivated by the fact that such regions relatively close to the star are probably more likely to experience collision-induced breakups of large objects. This is mainly because orbital velocities, and thus impact speeds, are higher closer to the central star, and also because inner disc regions are expected to be more densely populated. As a result, these regions should experience both more frequent and more violent (i.e., destructive) impacts. Note that this 6 AU distance also roughly corresponds to the best resolution obtained on resolved images of debris discs \citep[see for example][for the archetypal $\beta$ Pictoris system]{mill14}. Last but not least, 6\,AU is also the location where a high level of dustiness has been observed around HD172555, one of the best known systems with high luminosity excesses \citep{liss09,john12}.

As for the central star, we consider a typical A7V stellar type, motivated by the fact that the majority of imaged debris discs have been observed around early-type stars (see http://www.circumstellardisks.org). For estimating observed luminosities, we consider that the star+disc system is at a distance of 30 pc.

\begin{figure}
\includegraphics[scale=0.25]{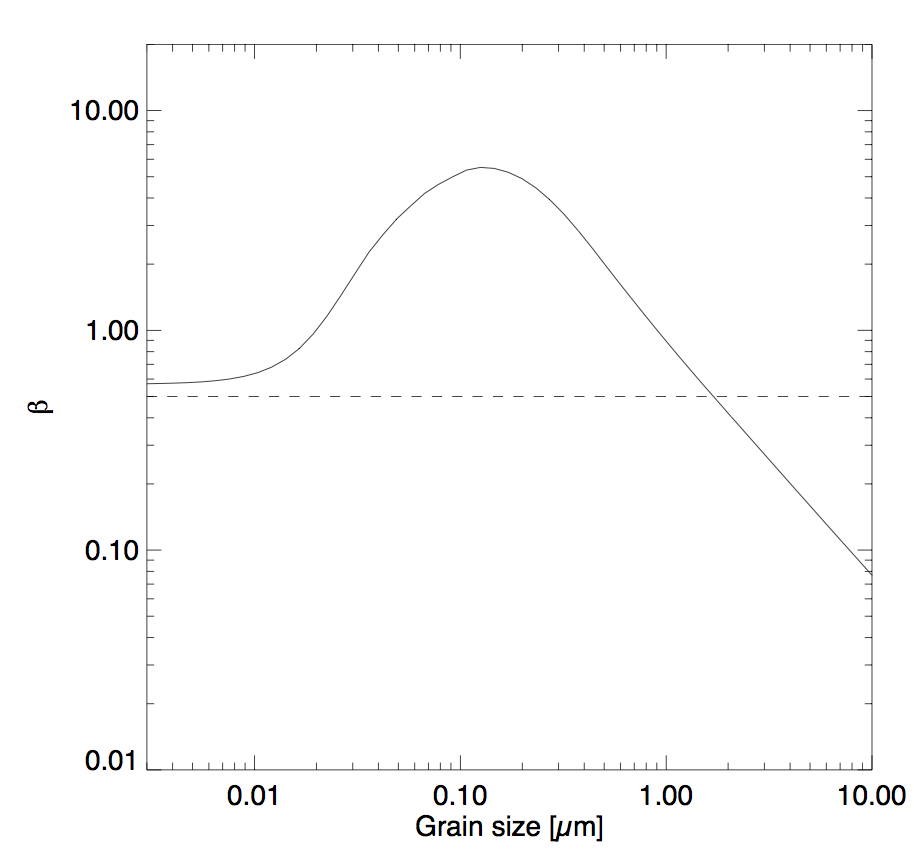}
\caption{$\beta$ values for astrosilicates around an A7V star. The dashed line shows the $\beta=0.5$ value.}
\label{betavalue}
\end{figure}

For the grain composition, crucial for the calculation of collision outcomes and for estimating their response to radiation pressure (value of $\beta$), but also for the production of synthetic images and SEDs with GRaTer, we consider generic astrosilicates \citep{drai03}.  Fig.~\ref{betavalue} displays the $\beta$ for the considered grain composition.

The differential size distribution of the initial fragments follows a steep power law in $dN/ds \propto s^{-3.8}$, corresponding to the crushing law expected for the outcome of violent collisions \citep{taka11,lein12}. The minimum size is taken to be around the blow-out size $s_\text{cut}$ induced by radiation pressure, i.e., $\simeq 1.8\mu$m for compact astrosilicates around an A7V star. We chose to discard smaller unbound grains because several test simulations have shown that, for the nominal set-up considered here, their contribution to the total surface density and luminosity is always negligible. The maximum size of the initial fragments is set at 1m.

The setup is summarized in Table~\ref{tabtest}.

\section{Results}\label{results}

We analyze the system's evolution using 5 types of outputs: smoothed 2-D surface density maps to characterize the spatial signature of the breakup's aftermath, disc-integrated evolution of the particle size distribution (PSD), optical depth profiles along the X-axis, disc-integrated luminosities and SEDs derived with GRaTer, as well as idealized synthetic images obtained with GRaTer and more realistic images simulated for the SPHERE/VLT and MIRI/JWST instruments.
In order to clearly identify the crucial role of collisions in the system's evolution, we also consider, as a comparison reference, a fiducial case without collisions, where only the dynamical evolution of the initially released fragments is considered. 

\begin{figure*}
\makebox[\textwidth]{
\includegraphics[scale=0.27]{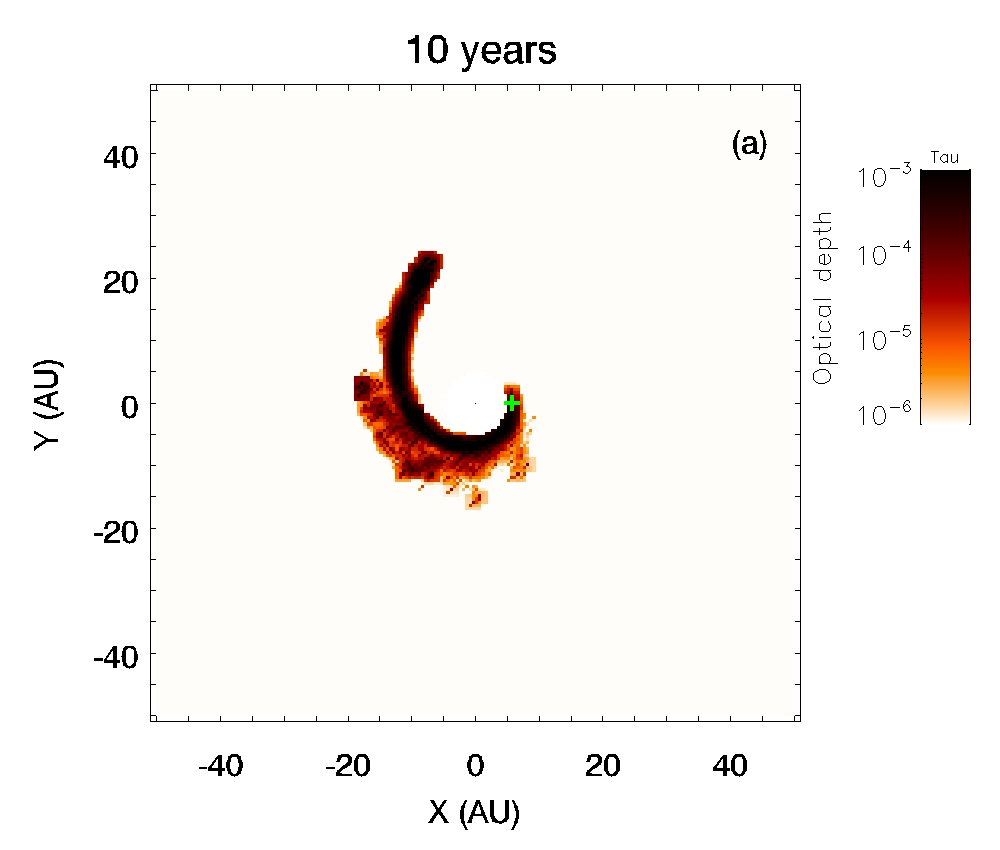}
\includegraphics[scale=0.27]{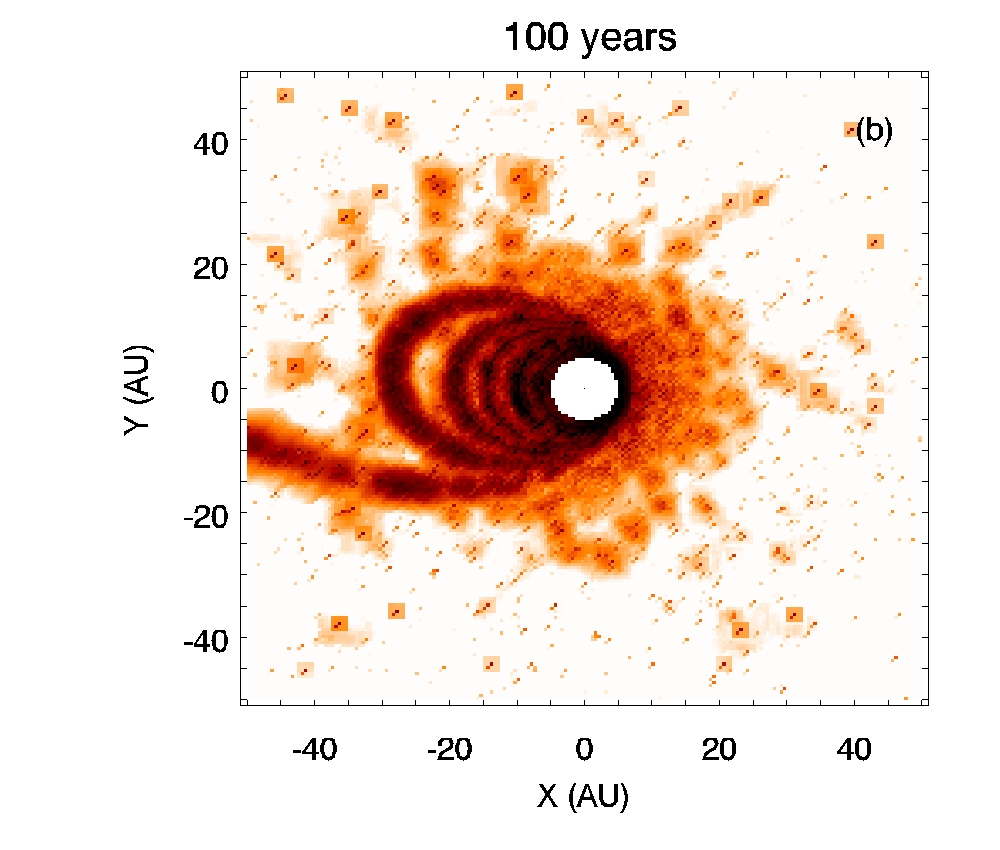}
}
\makebox[\textwidth]{
\includegraphics[scale=0.27]{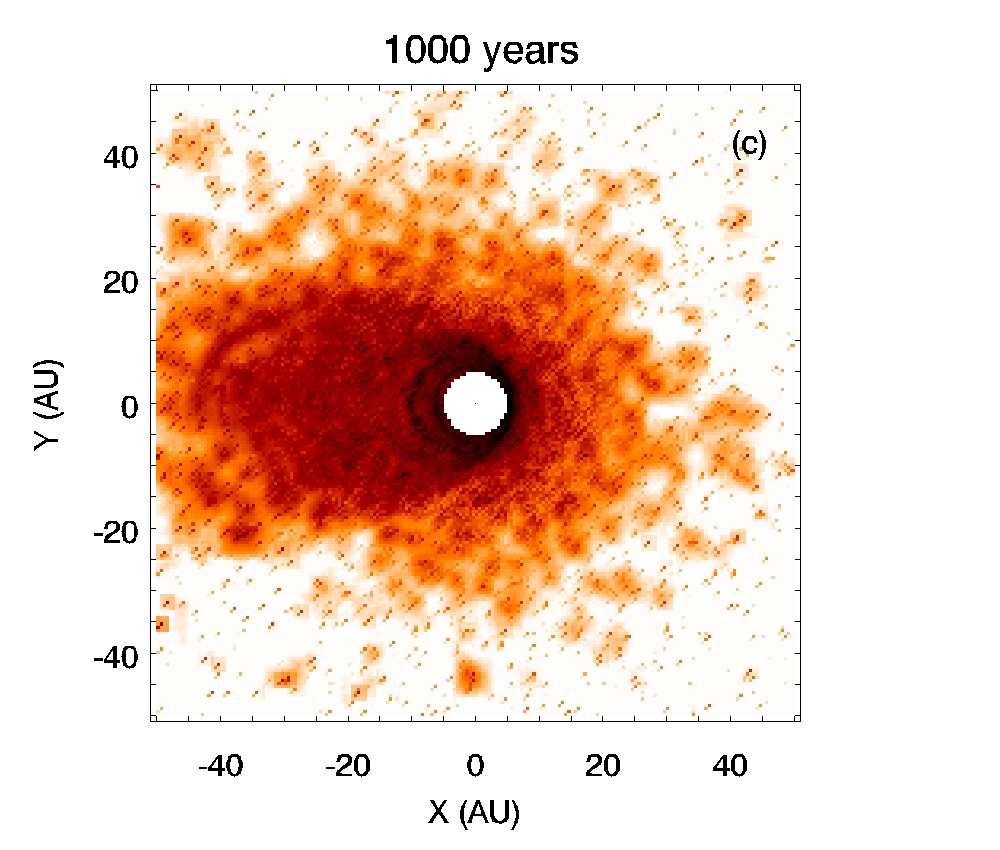}
\includegraphics[scale=0.27]{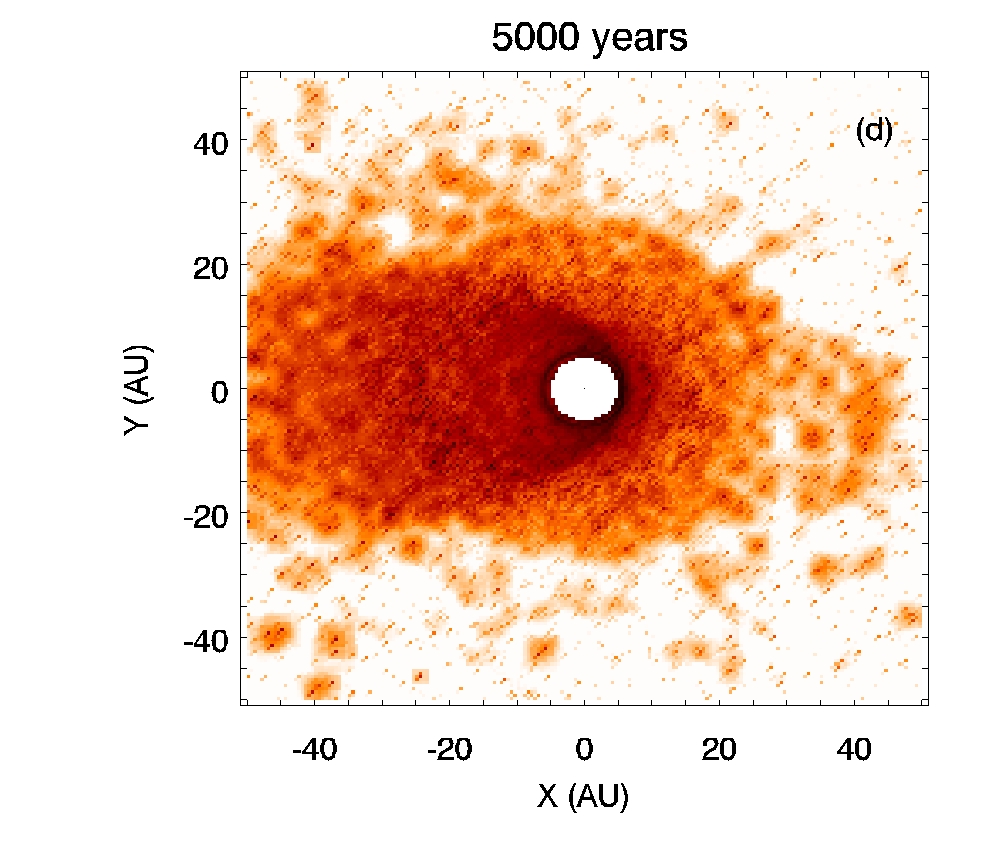}
}
\makebox[\textwidth]{
\includegraphics[scale=0.27]{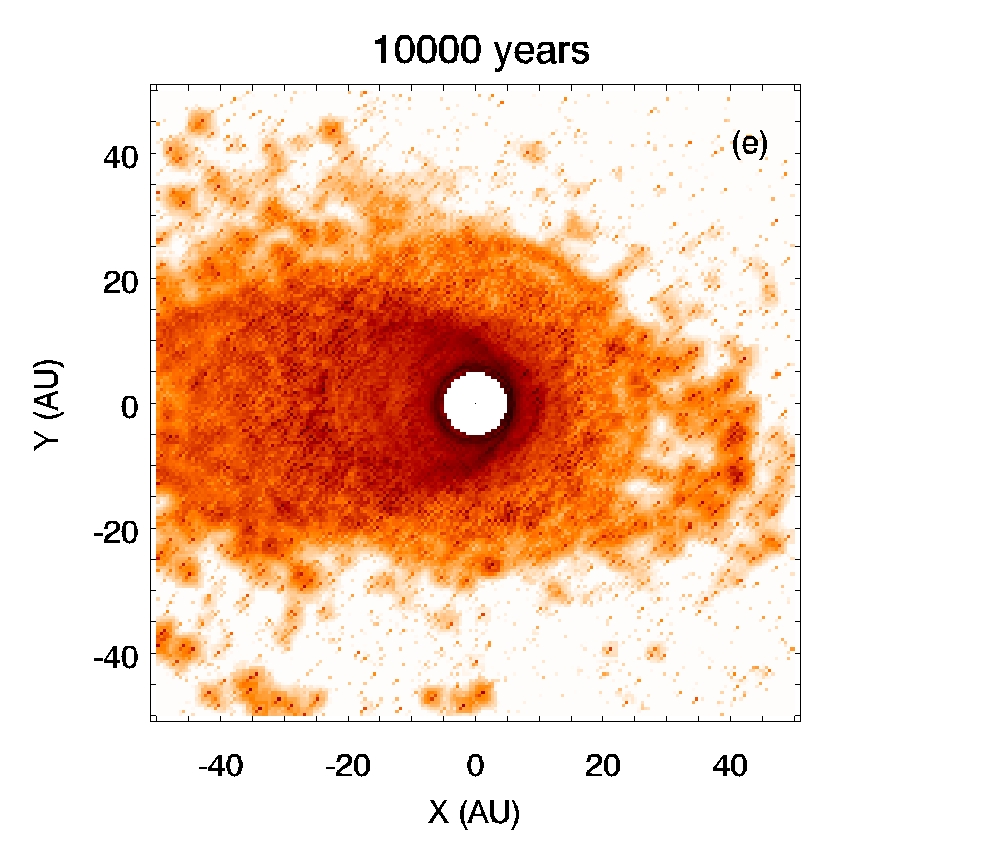}
\includegraphics[scale=0.27]{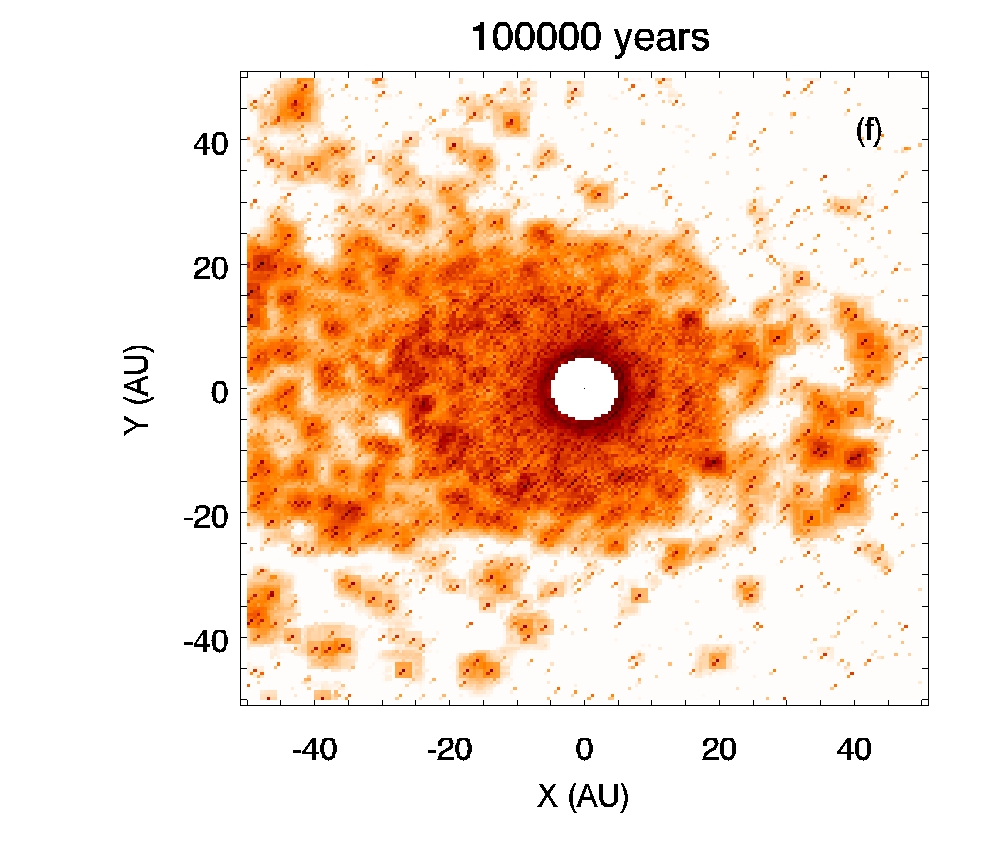}
}
\caption[]{Evolution of the system after the release of $10^{21}$kg of material at 6 AU from the central A7V star. 2-D map, in a non-rotating inertial frame, of the optical depth at different epochs after the initial breakup. The green cross on plot (a) is the location of the initial breakup. The color index goes from $10^{-3}$ (black) to $10^{-6}$ (light orange). }
\label{simuodgeneric}
\end{figure*}

\begin{figure*}
\makebox[\textwidth]{
\includegraphics[scale=0.27]{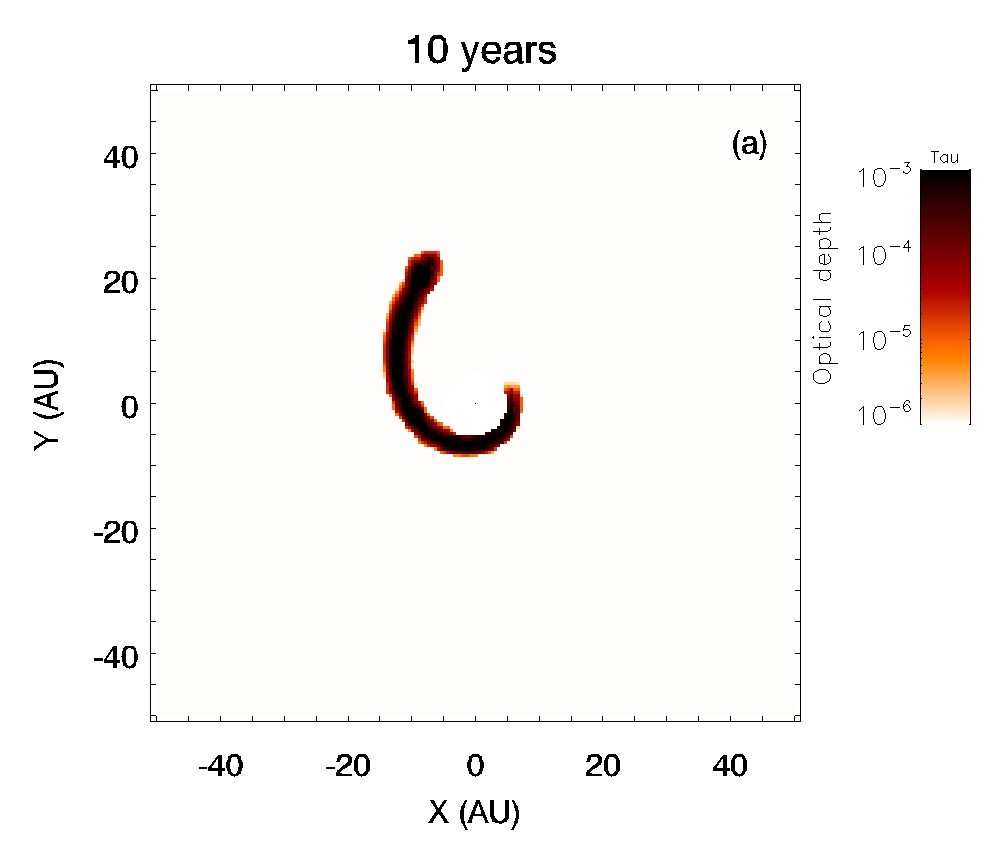}
\includegraphics[scale=0.27]{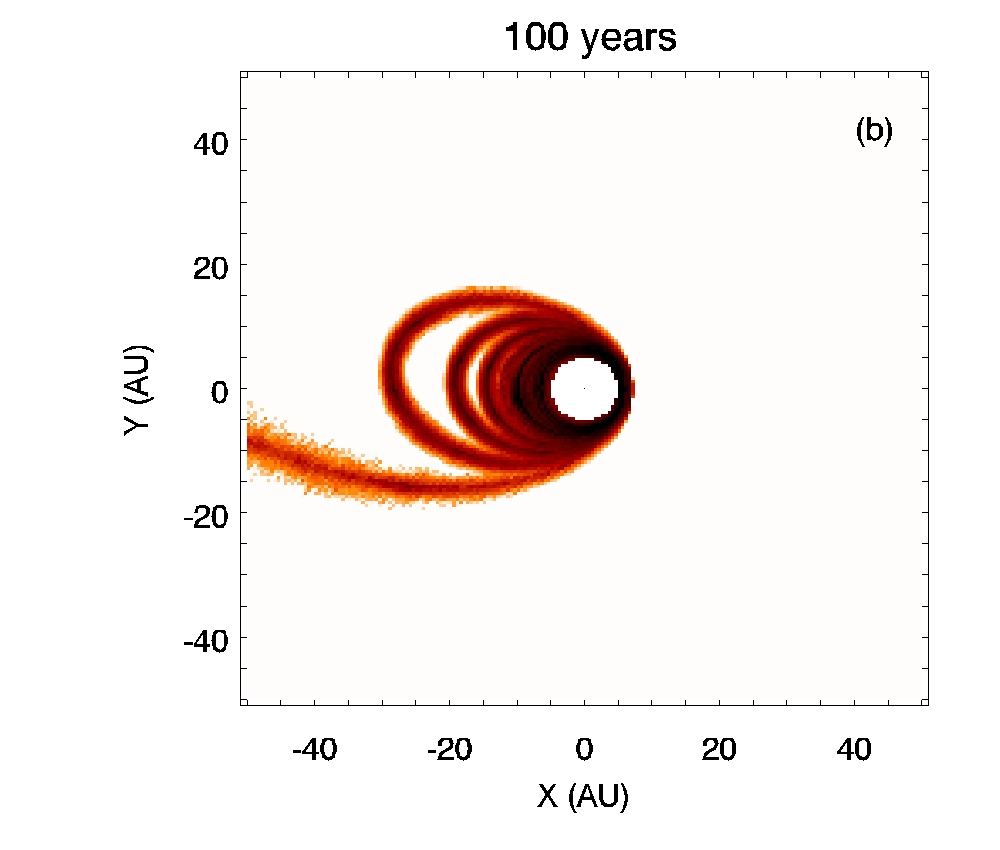}
}
\makebox[\textwidth]{
\includegraphics[scale=0.27]{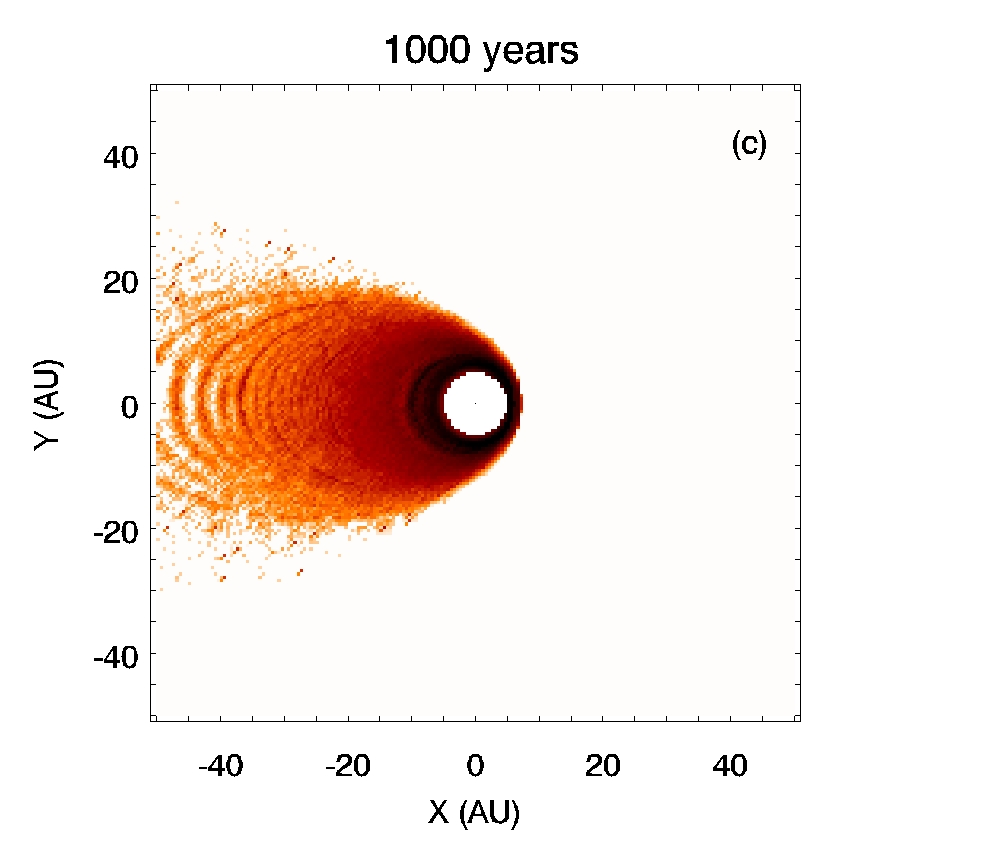}
\includegraphics[scale=0.27]{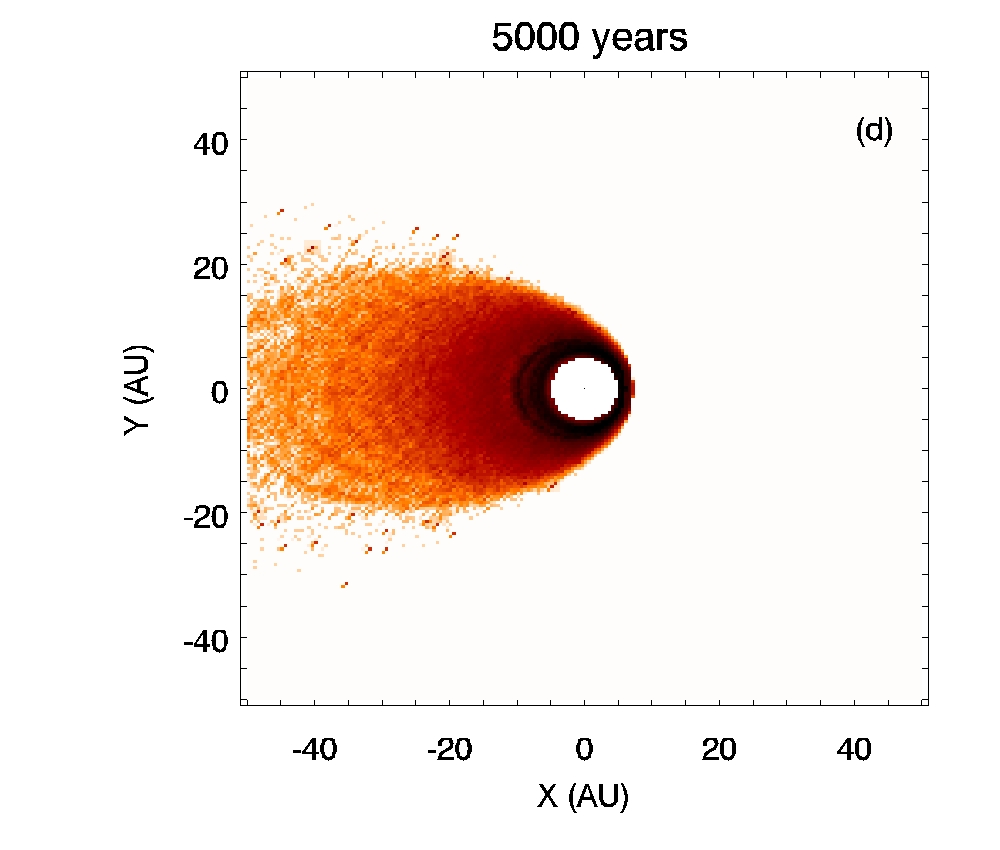}
}
\makebox[\textwidth]{
\includegraphics[scale=0.27]{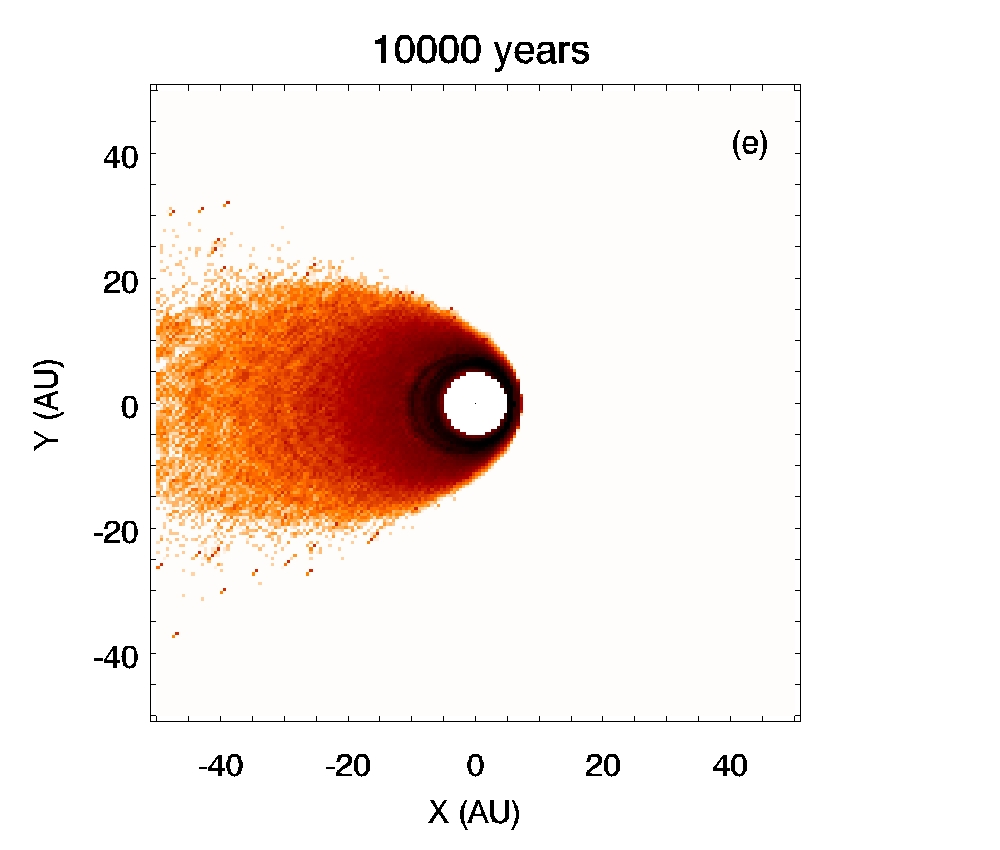}
\includegraphics[scale=0.27]{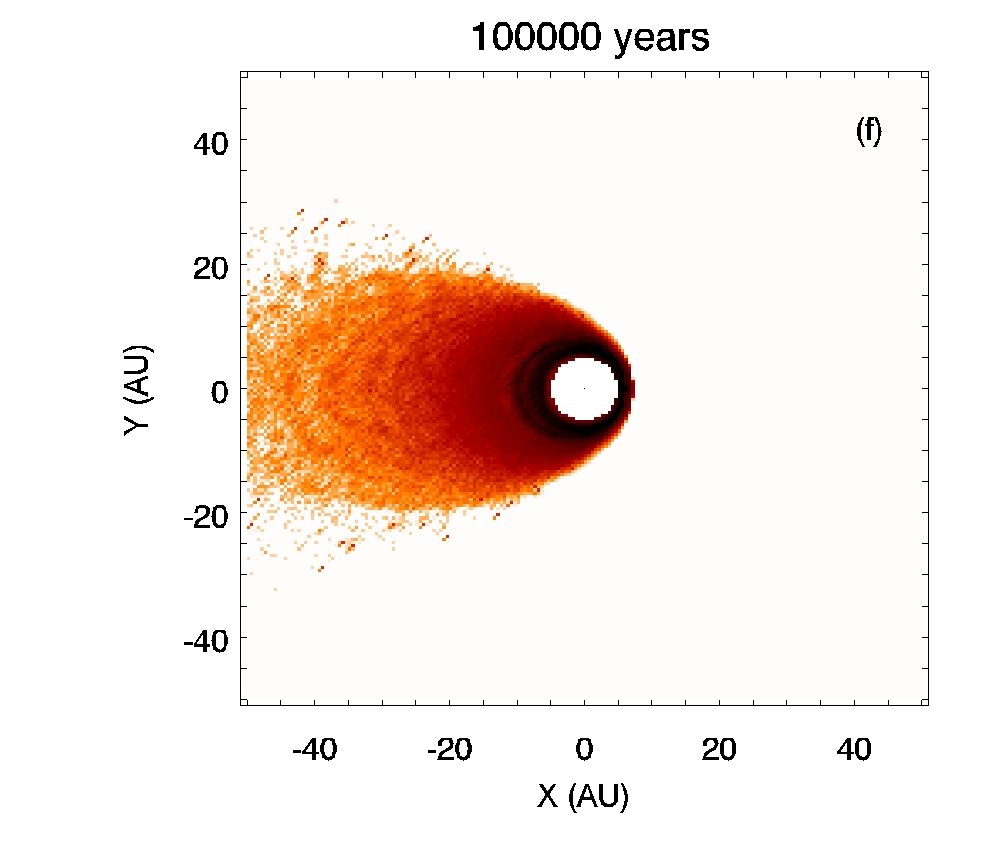}
}
\caption[]{Fiducial case without collisions, where only the dynamical evolution of the initial fragments is taken into account. }
\label{simdyn}
\end{figure*}

\subsection{Spatial Signature}\label{signature}

Fig.~\ref{simuodgeneric} displays the evolution of the system's surface density with time. This evolution can be schematically divided into 3 phases of unequal lengths.

\subsubsection{Transient ``spiral and ripples'' phase} \label{spiral}

In the immediate aftermath of the initial breakup, a one-armed spiral forms and propagates outwards (Fig.~\ref{simuodgeneric}a). This spiral corresponds to the peak luminosity of the system's post-breakup evolution (see Fig.~\ref{flux1}). Its outer parts consist mostly of small grains close to the blowout size, which are placed on highly-eccentric orbits by radiation pressure. This spiral structure rapidly fades out, in less than 100\,years, and morphs into elongated concentric ``ripples'', which become more and more tightly wound with time. These features arise because all released fragments' orbits have to pass through the initial release position at X=6 AU, Y=0. As a consequence, at a given time $t$, only particles having covered $n+0.5$ orbits will cut the X-axis on the left-hand side, that is only particles with $t_\text{orb}=t/(n+0.5)$. Since particles have different $t_\text{orb}$ because of their different $\beta$ values (which give them different $a$ and $e$), and since for any value of $t$ there is only a finite number of possible $n$ values, it follows that only particles from $n$ finite different size ranges cross the X-axis at any given time $t$, hence the appearance of $n$ ``ripples''. The timescale for setting up these ripple structures is thus typically a few orbital periods of the small grains populating them, i.e., a few times $t_\text{orb}=(1-\beta)/(1-2\beta)^{1.5} \times t_{\text{orb}(\beta=0)}$, where $t_{\text{orb}(\beta=0)} = 11$ years is the orbital period, at 6\,AU from the considered $1.84M_\odot$ star, for a large, non-radiation-pressure affected body. For the smallest grains considered in the runs, which have $\beta=0.44$, we thus get a typical timescale of a few $t_\text{orb}\sim 150\,$years. Note that such ripples have also been identified by \citet{jack12} and \citet{jack14}, but in a different context where different $t_\text{orb}$ values are not due to different $\beta$ values (i.e., grain sizes) but to the very large initial spread in post-impact ejection velocities. 

As time goes by, however, these ripple features fade away and become undetectable after $\sim 1000\,$years (Fig.~\ref{simuodgeneric}c). This is firstly because the ripples become more and more wound up so that they tend to merge into each other. In addition, collisions amongst the debris, both through the feedback they have on the dynamics and also the production of new particles from parent grains within the ripples, greatly accelerate this smearing out of the ripples. This is clearly illustrated when looking at the evolution of the fiducial case without collisions (Fig.~\ref{simdyn}), for which ripples are still clearly visible after 1000\,years, and only vanish after $\sim10^{4}\,$years.

\subsubsection{Asymmetric Disc} \label{asymd}

After the disappearance of the ripple features at $\sim 1000\,$years, the system enters a second, more long-lived phase where it assumes the shape of an asymmetric eccentric disc (Fig.~\ref{simuodgeneric}c,d,e). This elongated shape is due to the fact that, during this period, most grains still have their orbits passing by a point close to the initial release location, and that the disc's geometrical cross section is dominated by small grains, close to the blowout limit, which have high-$e$ orbits because of radiation pressure.
Note, however, that the bulk of the system's $mass$, which is contained in the biggest particles, is located in a nearly-circular ring passing by the release point at 6 AU. This ring forms due to Keplerian shear over a few dynamical periods. Its width is set by the initial velocity dispersion of the post-release fragments. This ring logically also corresponds to the region where most of the collisional activity takes place. 

\begin{figure*}
\makebox[\textwidth]{
\includegraphics[scale=0.35]{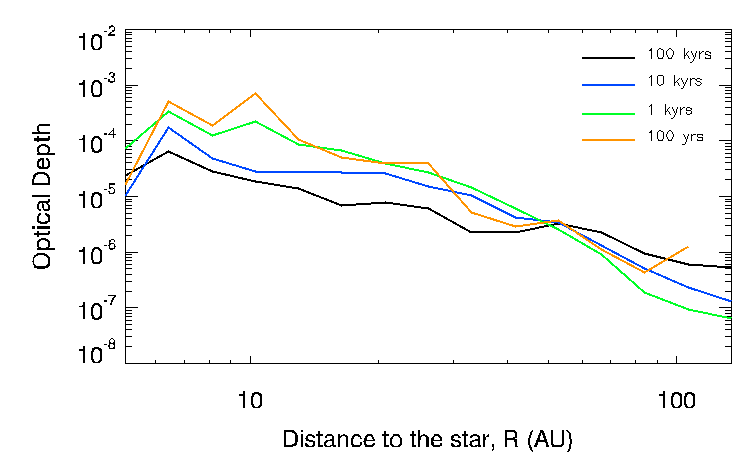}
\includegraphics[scale=0.35]{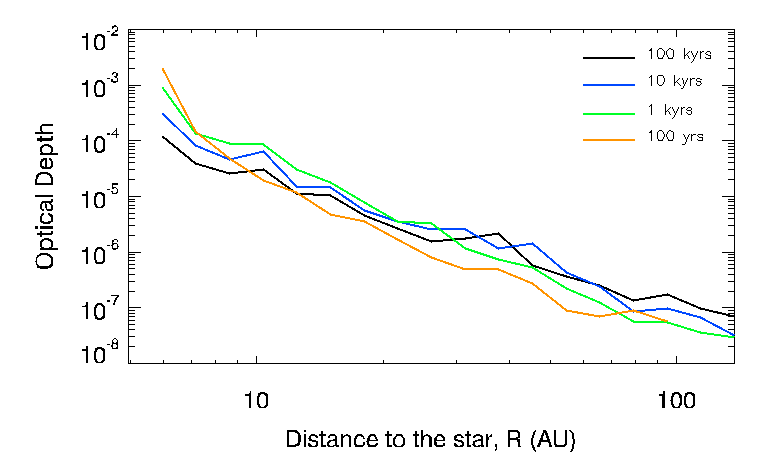}
}
\caption[]{Cut along the X-axis. Vertical optical depth profile at different epochs after the initial breakup (Note the log scale on both axis). Left: Cut along the left-hand side (X < 0), right: Cut along the right-hand side (X>0)}
\label{simucutx1}
\end{figure*}

During this asymmetric disc phase, there is a pronounced brightness asymmetry between the left and right-hand sides, as is clearly illustrated by Fig.~\ref{simucutx1} showing cuts along the X-axis. This asymmetry is due to the over-density of material created by the clustering of orbits close to the release point on the right-hand side. It is, moreover, reinforced by the increased collisional activity, and thus dust production, that takes place in this over-dense region of the inner ring. As a consequence, in the inner regions (6\,AU), the right hand side appears much brighter than the opposite one. In the outer regions (beyond 10-20\,AU), however, the reverse is true: the left hand side is brighter. This is because it is densely populated by small high-$\beta$ grains, produced at the breakup point and passing close to their apastron, whereas such small grains are much less abundant on the right hand side because they originate from the more tenuous left hand side of the inner ring.

\subsubsection{Progressive symmetrization}\label{progsym}

As time goes by, however, the asymmetric elongated disc structure progressively fades out. Figs~\ref{simuodgeneric}d,e,f clearly show that the initially tenuous outer regions of the right-hand side progressively fill up with matter. This matter consists mostly of small grains, placed on high-$e$ orbits by radiation pressure, which have been produced by second (or more) generation collisions in regions that are no longer that of the initial breakup. One of the main production sources is the aforementioned inner ring made of all the biggest initially released fragments. The grains produced in this ring can have any orbital orientation depending on which part of the ring they have been produced. Of course, the collisional activity is, at least in the early phases, higher at the part of the ring corresponding to the breakup location, because, in addition to the large fragments, it is also populated by smaller grains passing by their release point. However, as time passes, the density contrast within the ring decreases, and so does the difference in collisional activity between the breakup location and the rest of the ring (see Fig.~\ref{simucutx1}), so that small dust produced in the inner ring tend to symmetrize the system.

\begin{figure}
\includegraphics[scale=0.36]{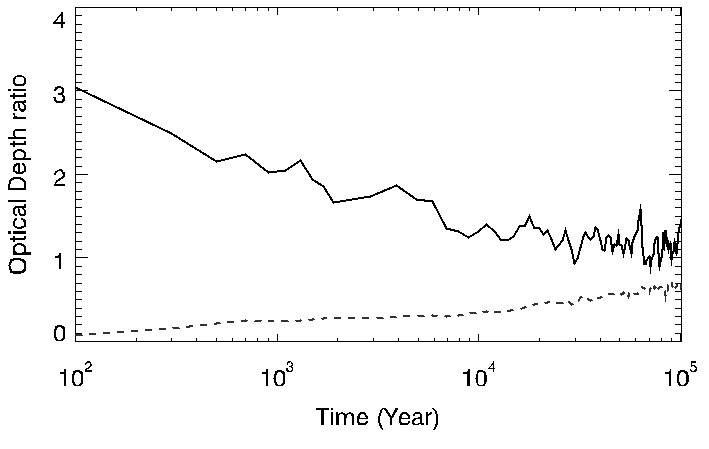}
\caption{Evolution of the system's left-right asymmetry, as quantified by the right-to-left cross section ratio within the inner ring (top) and in the region beyond it (bottom). }
\label{asym}
\end{figure}

To quantify the progressive decrease of the system's asymmetry, we plot on Fig.~\ref{asym} the relative contrast between the right and left-hand side luminosities, both in the inner ring and beyond it. We see that the asymmetries are resorbed by collisions after $t\sim5\times10^{4}$ years in the inner ring while the resorption timescale in the outer regions is $\sim10^{5}$ years. This is in agreement with Fig.~\ref{simuodgeneric}f showing an almost axisymmetric disc at $t\sim10^{5}$ years.

These timescales can be understood in analytical terms, as it comes down to estimating the timescale for collisions to even out the densities within the inner ring. To a first order, this timescale corresponds to the time it takes to collisionally reprocess all the material available at the initial collisional release point, which is approximately the collisional lifetime $t_\text{col}$ of the largest objects in the collisional cascade at this point. A rough estimate of $t_\text{col}$ can be obtained using the following formula, derived by \citet{lohn08} for a narrow ring of radial distance $r$ and width $dr$:
\begin{equation}
t_\text{col} = \frac{4 \pi}{\sigma_\text{tot}} \left( \frac{s}{s_\text{min}} \right)^{-q_p-3}\frac{r^{5/2} \text{d}r}{\sqrt{GM_\star}}\frac{I}{f(e,I)G_l(q_p,s)},
\label{equlohn}
\end{equation}
where $\sigma_\text{tot}$ is the initial total cross-sectional area, $q_p$ the slope of the primordial size distribution \footnote{Note that in L\"{o}hne et al.'s original formula $q_p$ is the index of the mass distribution, we have adapted the equation to the present notation.}, $s_\text{min}$ the blow-out size, $e$ is the mean eccentricity, $I$ the inclination, $f(e,I)=\sqrt{5/4e^2+I^2}$, and the function $G_l$ is defined in equation (24) of \citet{lohn08}.
Note that L\"{o}hne's formula is valid for a uniform disc, which is not the case here. To roughly correct for this, we increase $\sigma_\text{tot}$ by a factor 2, which is an approximate average of the right/left density contrast (see Fig.~\ref{asym}). 
With the present set-up ($M_\star=1.84M_\odot$, $q_p=3.8$, $r=6$\,AU, $\text{d}r=1\,$AU and $e=2I=0.037$), we then get $t_\text{col}\sim 2.4 \times 10^5\,$years, which is relatively close to what is found in our simulation.
Note, however, that Equ.~\ref{equlohn} can only give an order-of-magnitude estimate, because, in addition to ignoring the asymmetric character of the ring, it also neglects the crucial role of cratering impacts (only fragmenting collisions are taken into account in this equation)

\subsection{Particle Size Distribution}\label{PSD}

\begin{figure}
\includegraphics[scale=0.35]{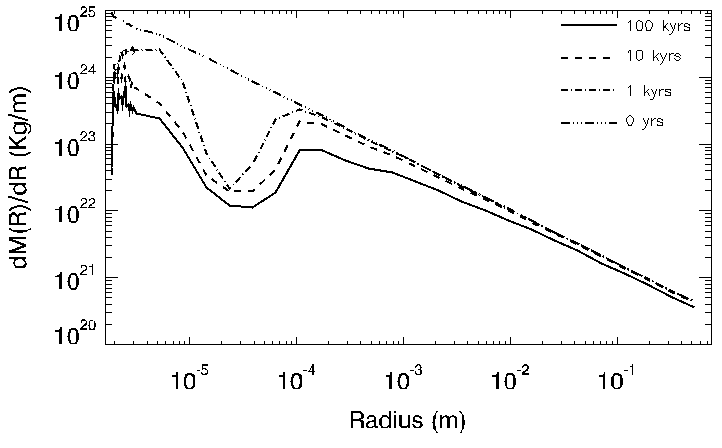}
\caption{Evolution of the differential mass distribution $dM/dR$ within the inner main ring (at 6AU) at 4 different epochs after the initial breakup.}
\label{simumdgeneric}
\end{figure}

Fig.~\ref{simumdgeneric} shows the evolution of the particle size distribution (displayed in the form of the differential mass distribution) within the main ``inner ring'' at $\sim$ 6 AU (where the bulk of the mass is located, see previous section). As can be clearly seen, the PSD undergoes important changes after the initial release. 

Within this inner ring, the initial PSD, corresponding to the ``crushing law'' of the initial fragments, is progressively relaxed due to collisional processing. The system evolves towards a collisional steady-state resembling those obtained by previous statistical codes for collisional debris rings \citep[see for example][]{theb08}. At the lower end of the PSD, one important feature is a ``wavy'' pattern extending up to $\sim100\,s_\text{cut}$ \citep[see][ for a thorough description of this feature]{theb07}, which develops on a timescale of the order of $10^{4}$ years.
For larger particle sizes, the PSD evolves more slowly. It progressively departs from the initial crushing law in $s^{-3.8}$ and tends towards an equilibrium law in $s^{-3.66}$, which is the value expected as a consequence of the size dependence of the critical specific energy $Q^*$ assumed here (see Section \ref{model})\footnote{Note that this flattening of the PSD also affects the small size regime and is clearly visible despite the pronounced wavy structure. This flattening of the PSD tends to lower the amount of small particles next to the cut-off and, as a consequence, the total optical depth.}. 
This steady-state regime in $s^{-3.66}$ progressively works its way up towards larger particles, mainly because larger objects have longer collisional lifetimes. At $t=10^{5}$ years, the steady-state PSD has reached the maximum size of our size distribution at $s\sim1$\,m. 

Note that these relaxation timescales within the inner ring, that is, $\sim10^{4}$ years to reach steady state in the small-grains domain and $\sim10^{5}$ years in the larger bodies domain, are relatively independent of the initial crushing law for the released fragments, as could be verified with test runs exploring steeper and shallower initial slopes. 

\subsection{Detectability}\label{detect}

\begin{figure*}
\makebox[\textwidth]{
\includegraphics[scale=0.25]{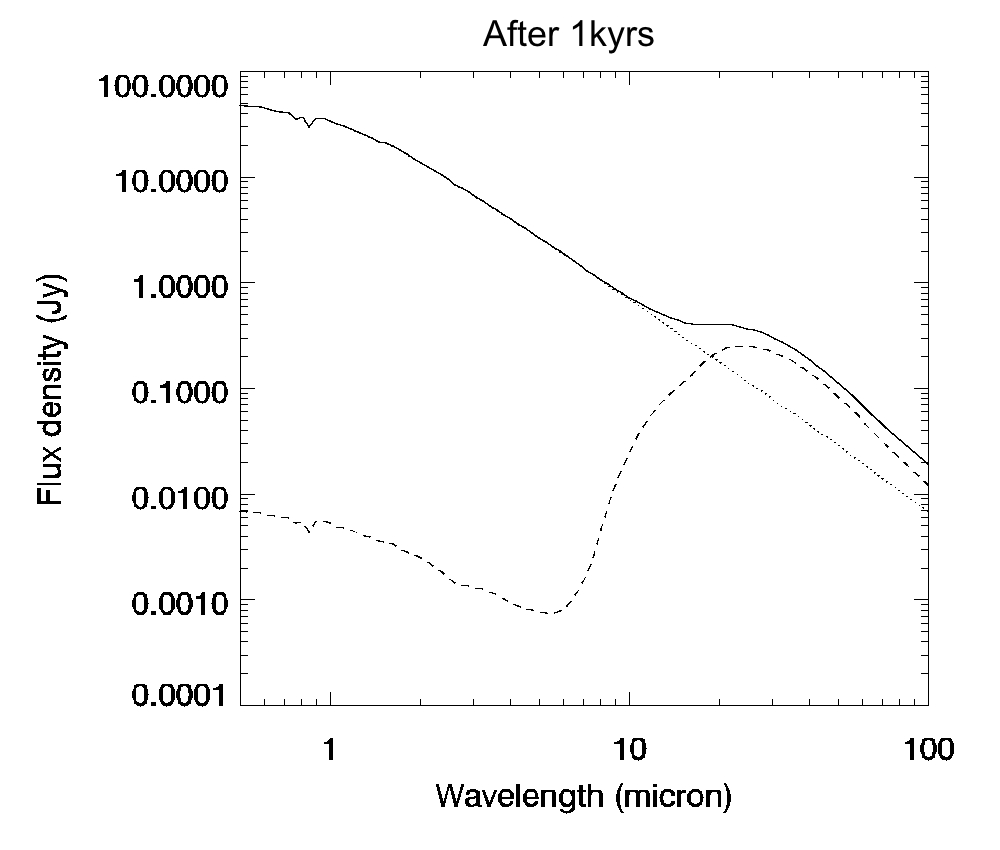}
\includegraphics[scale=0.25]{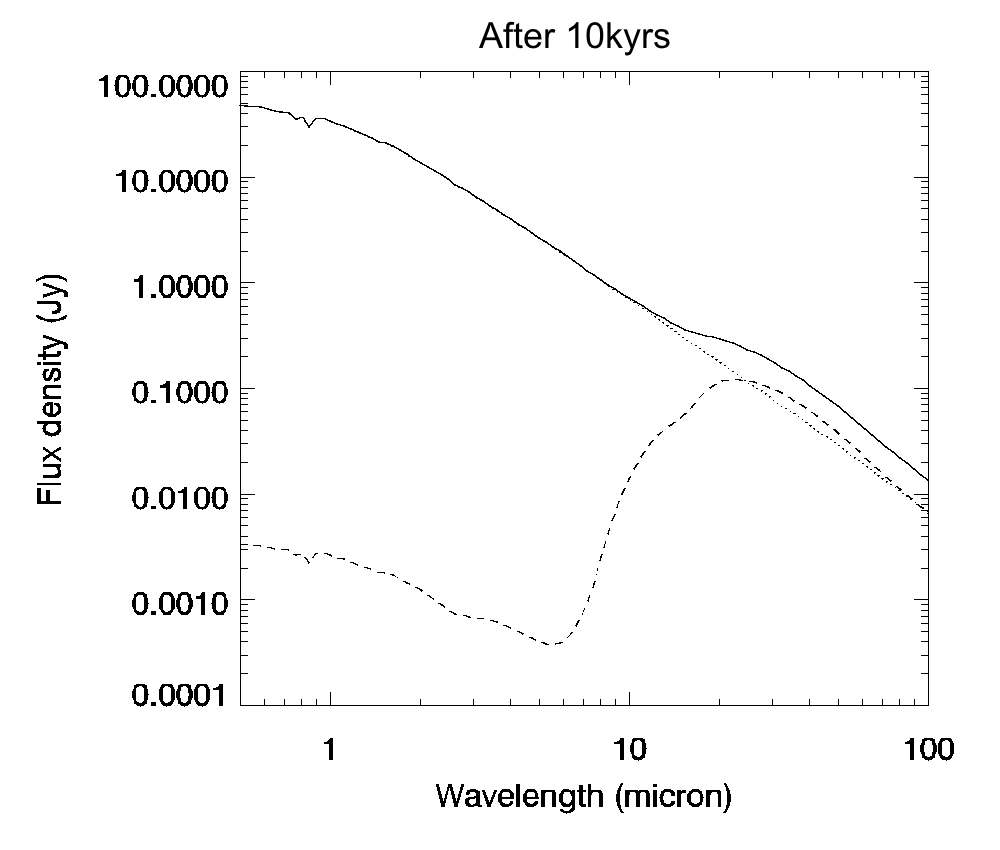}}
\caption{SED of the integrated system, as estimated with the GRaTer package at 1 (left) and 10 kyrs (right) after the breakup. The full line represents the total (stellar photosphere + disc) luminosity, while the dashed line represents the sole contribution of the dust disc and the dotted one is that of the stellar photosphere. The scattered light contribution to the disc SED, which dominates over thermal emission at short wavelengths, has been calculated assuming isotropic scattering.}\label{SED}
\end{figure*}

\begin{figure}
\includegraphics[scale=0.33]{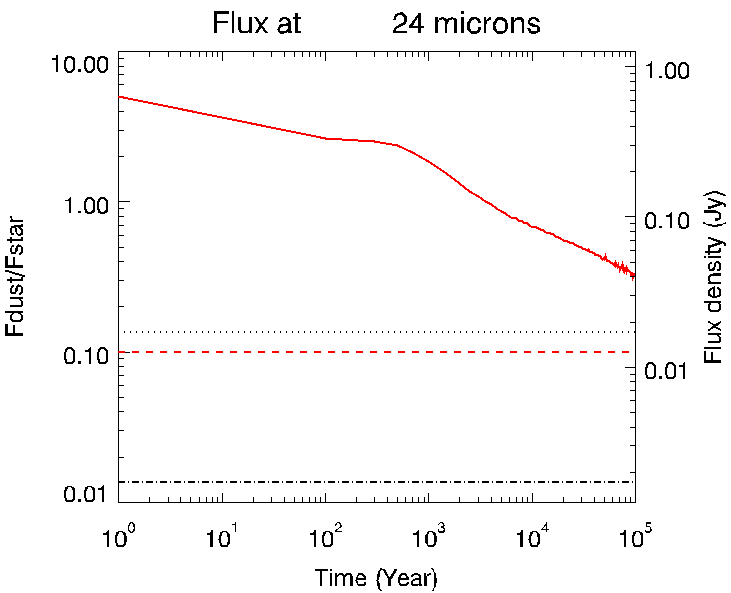}
\caption{Evolution of the disc-integrated flux at 24 $\mu$m, as computed with GRaTer (red solid line). The red dashed line marks a dust-to-star flux ratio equal to 0.1, taken as our detectability criteria. The dotted and dash-dotted black lines give the maximum possible luminosity, at 1Myrs and 10Myrs, for a hypothetical collisional cascade at steady-state (see text for details). The X-axis indicates the time after the breakup in years. The right-hand side Y-axis indicates the absolute flux in Jy, while the left-hand side axis displays the ratio of the disc flux to that of the stellar photosphere.}
\label{flux1}
\end{figure}

\begin{figure*}
\makebox[\textwidth]{
\includegraphics[scale=0.27]{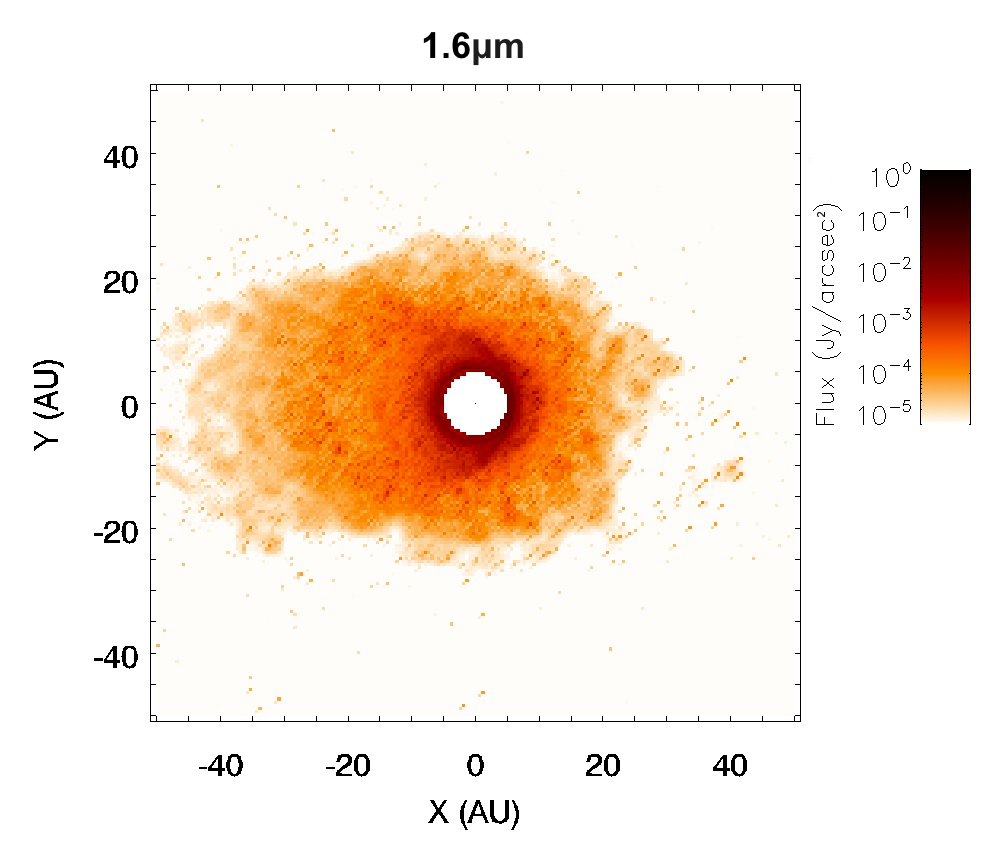}
\includegraphics[scale=0.27]{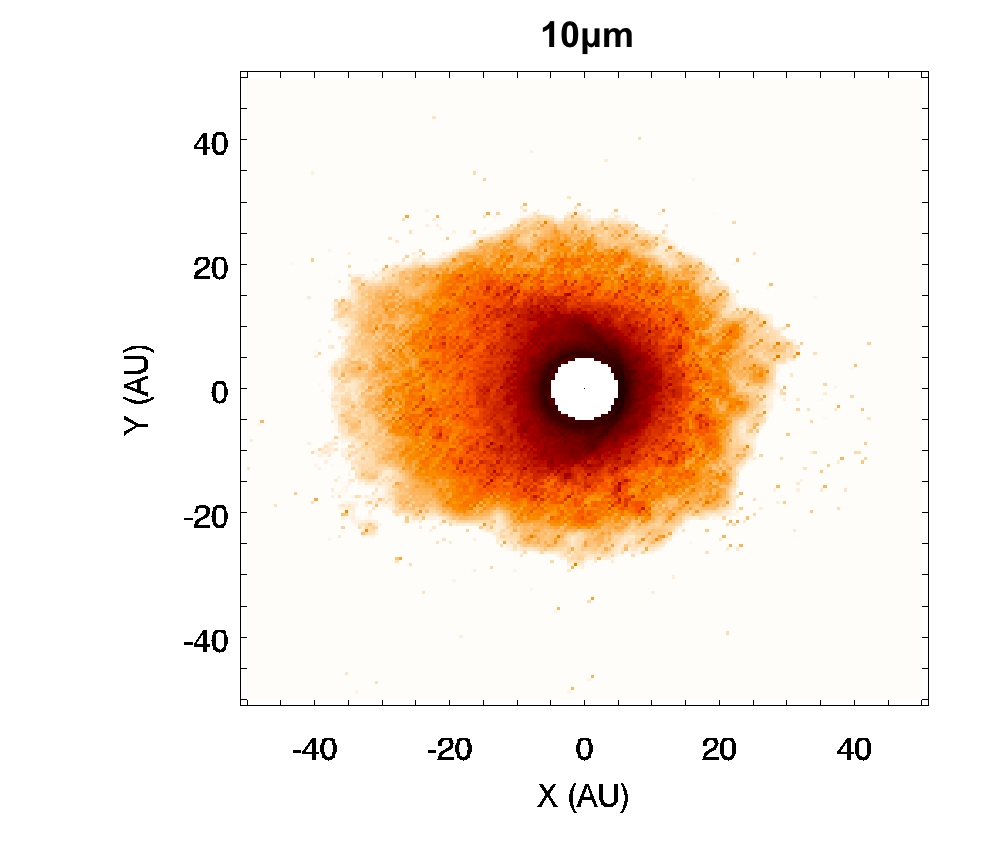}
}
\makebox[\textwidth]{
\includegraphics[scale=0.27]{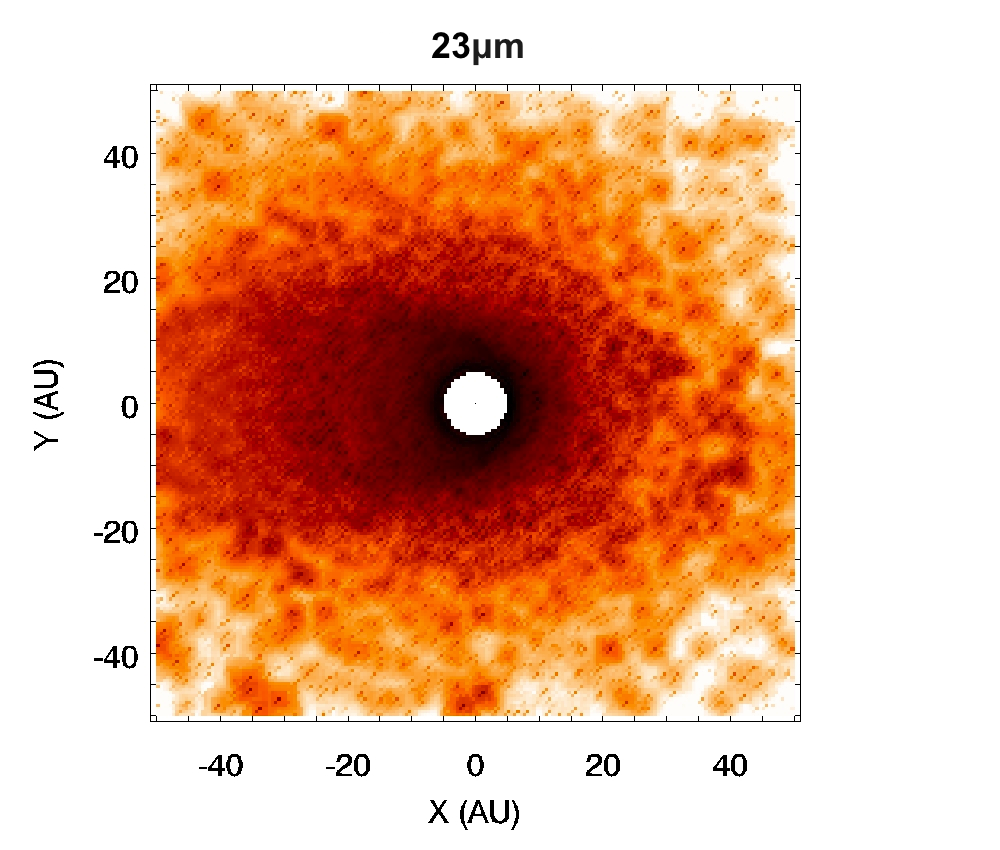}
\includegraphics[scale=0.27]{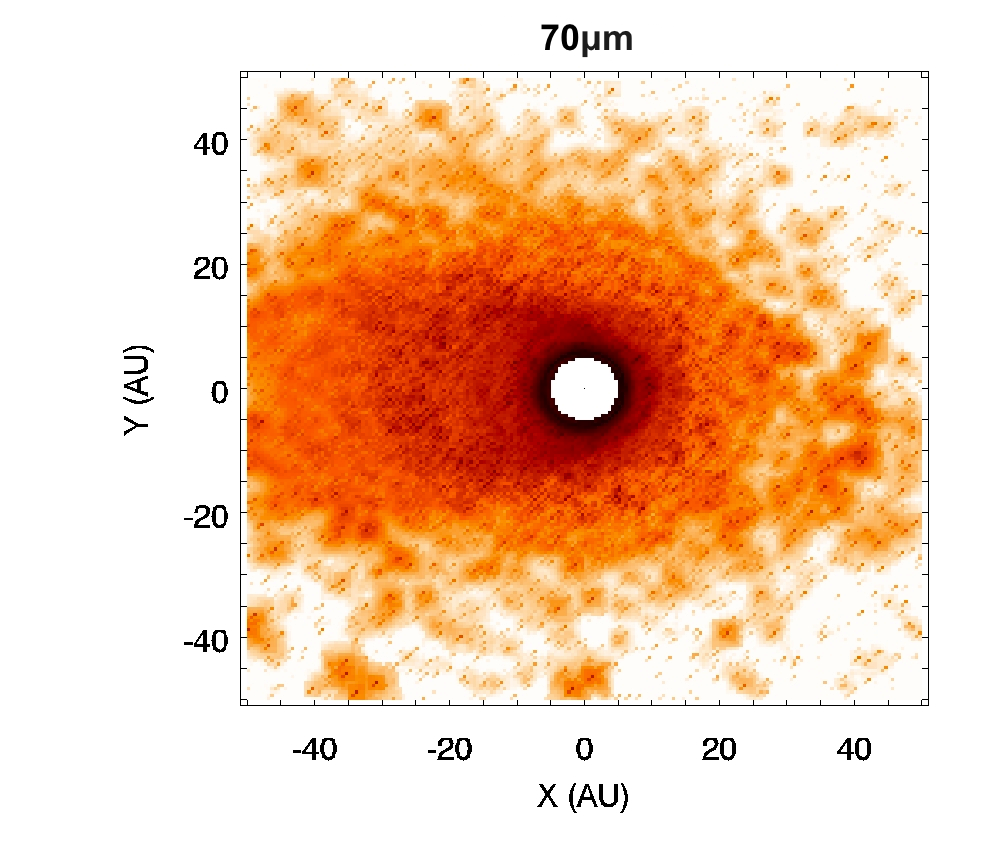}
}
\caption[]{Idealized synthetic images of the system during its asymmetric disc phase (at t=$10^4$ years), produced with the GRaTer package. We consider a system observed at 30pc, at 4 wavelengths: 1.6$\mu$m (upper left panel), 10$\mu$m (upper right), 23$\mu$m (lower left) and 70$\mu$m (lower right). The color-scale is the same for all 4 images. }
\label{synthimage}
\end{figure*}

Fig.~\ref{SED} shows the star+disc integrated SED, at two different epochs, as computed with the GRaTer package. The excess due to the post-breakup dust appears clearly in the mid-IR domain, peaking around $25\mu$m. This is confirmed by the synthetic images obtained with the GRaTer package at different wavelengths, showing that the dust disc is at its brightest in this $\lambda \sim 25 \mu$m domain (see Fig.~\ref{synthimage}).

In order to estimate the detectability of this flux excess, we plot on Fig.~\ref{flux1} the evolution of the disc-integrated flux at $24\mu$m, as observed at a 30pc distance, as well as the disc-to-stellar-photosphere flux ratio at this wavelength. As a typical criterion for detectability, we take as a reference the performance of the Spitzer/MIPS instrument at 24 microns. We consider that the disc-to-star flux ratio should exceed 10\% for a 3 sigma, or larger, detection of an excess above the stellar photosphere. This ratio assumes a 2\% (1 sigma) calibration uncertainty of MIPS24 \citep{enge07}, and a 1 to 2\% (1 sigma) uncertainty on the predicted stellar flux at 24 microns. We also consider a typical absolute flux sensitivity of 1mJy at 24 microns.

As can be seen, for our nominal case, the disc signature remains detectable, at $24\mu$m, for the whole duration of the simulation ($10^{5}$ years). After an initial transient period of slightly decreasing flux, it reaches $F_\text{dust}/F_\text{star} \sim 3$ after $\sim500\,$ years. Beyond 500 years, the flux decreases with time following an approximate $t^{-0.3}$ power law. We note that this slope is in agreement with the theoretical decrease in $\sim t^{-0.31}$ of the dust mass predicted, for bodies in the strength regime, by Equ.~43 of \citet{lohn08} when using the parameters assumed in our simulations ($q_p=1.9333$ and $q_s=1.8866$) \footnote{The often invoked asymptotical decrease in $t^{-1}$ is only valid at times greatly exceeding the collisional lifetime of the largest bodies in the collisional cascade \citep[e.g.][]{lohn08}.}. 
Since $F_\text{dust}/F_\text{star}$ follows this approximate $t^{-0.3}$ decrease, it is easy to extrapolate the Fig.~\ref{flux1} curve to later times. The extrapolated time at which the system reaches the critical $F_\text{dust}/F_\text{star} \sim 0.1$ value is then $t_\mathrm{detect} \sim 10^6$ years, which gives the approximate duration of the detectability phase.

We note that $t_\mathrm{detect}$ exceeds the duration of the asymmetric-disc phase, which is $\sim 10^5$ years for the nominal set-up considered here (see Fig.~\ref{asym} and Sect.~\ref{signature}). This means that, beyond $\sim 10^5$ years after the initial impact, the system will be detectable in photometry but should have the aspect of an axisymmetric disc in resolved images. For the first $10^5$ years (and this is probably a lower limit, see Sect.~\ref{limit}), however, the asymmetric signature of the massive breakup should be clearly visible $if$ the system is observed with an instrument that can resolve it (see next section), as is clearly illustrated in the synthetic images obtained with GRaTer at 4 different wavelengths (Fig.~\ref{synthimage}). 

Another important result is that, in the inner disc regions ($\sim 6$ AU) considered here, the flux excess due to one massive impact greatly exceeds that of a debris disc at steady-state, i.e., a disc whose luminosity is due to a ``standard'' collisional cascade. We verify this by plotting in Fig.~\ref{flux1} the maximum possible luminosity, at 2 different ages (1 and 10 Myrs after the equilibrium collisional cascade phase is reached), expected for such a steady-state disc. This luminosity is estimated using the equation of \citet{wyat07}, assuming a 1\,AU wide disc centered at 6\,AU \citep[all other parameters being the same as those assumed by][]{wyat07}\footnote{Note, however, that this equation should be used with caution, as it strongly depends on the index assumed for the PSD, as well as on poorly constrained parameters, such as the orbital eccentricities of disc particles. It is used here as a convenient order of magnitude indicator. From the fractional luminosity given by this formula, we estimate the flux at 24$\mu$m assuming that the disc SED peaks at this wavelength.}. As can be clearly seen, our post massive-breakup disc is, for the whole duration of the simulation, at least one order of magnitude brighter than even a very young steady-state disc only 1 Myrs after its collisional cascade phase has begun.

\subsection{Imaging with SPHERE and MIRI} \label{images}

\begin{figure*}
\begin{center}
\includegraphics[scale=0.6]{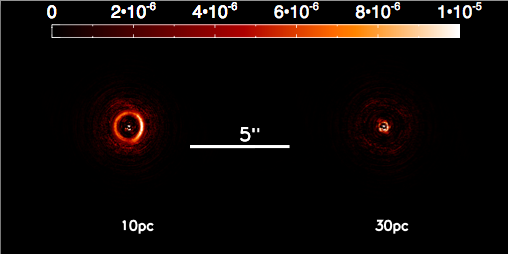}
\caption[]{Synthetic images at 1.6$\mu$m for the SPHERE/VLT instrument, for the system in its ``asymmetric disc'' phase ($10^{4}$ years), at 10 (left) and 30\,pc (right), obtained with the reference star subtraction method (see text for details). The color scale gives the flux ratio with respect to the brightest pixel in the point spread function (PSF).}
\label{SPHEREimage}
\end{center}
\end{figure*}

\begin{figure*}
\begin{center}
\includegraphics[scale=0.6]{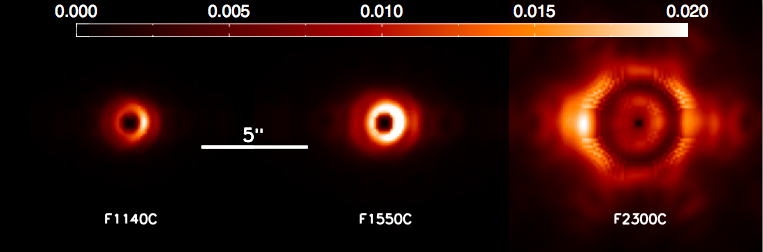}\\
\includegraphics[scale=0.6]{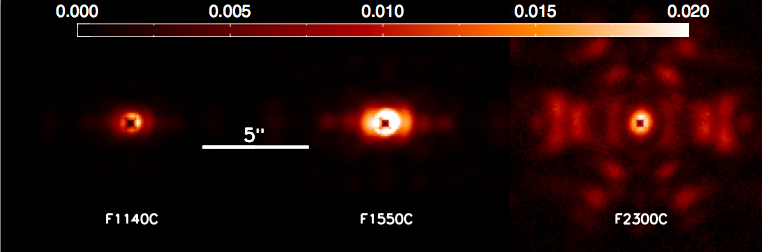}
\caption[]{Synthetic images, for the system in its ``asymmetric disc'' phase ($10^{4}$years), at 10 (top) and 30\,pc (bottom), with the MIRI/JWST instrument with different filters at 11.4, 15.5 and 23$\mu$m (from left to right) obtained with the reference star subtraction method (see text for details). The color scale gives the flux ratio with respect to the brightest pixel in the PSF.}
\label{MIRIimage}
\end{center}
\end{figure*}

\subsubsection{Method}

The images obtained with GRaTer in Fig.~\ref{synthimage} are idealized cases that do not take into account the limitations inherent to observation instruments. We decided to go a step further and test the observability of the collision-induced discs with the expected performance of two instruments: SPHERE, the near IR extreme AO-coronagraphic facility, now in commissioning phase at the VLT \citep{beuz08}, and MIRI, the mid-IR instrument of JWST \citep{wrig10} also including coronagraphs for exoplanets and disc observations. We here compare the intensity level of the synthetic disc images to the residual starlight after processing the data, determined from simulations. 

We consider our star-disc system at two distances, 10\,pc and 30\,pc. The dust distribution obtained from the simulations is converted to flux ratio at 1.6$\mu m$ (SPHERE) and  11.40, 15.50 and 23 $\mu m$ for MIRI, and sampled at the appropriate pixel scales.

To obtain the near-IR images we use the simulation results presented in \citet{bocc08} as well as the very first contrast estimations of SPHERE in the lab \citep{lang13}.
The default observing mode of SPHERE for exoplanet detection \citep[angular differential imaging, ][]{maro06}, is inappropriate in that case as it causes self-subtraction, which can be problematic when dealing with extended objects like circumstellar discs \citep{bocc13}.
Instead, we used the method of reference star subtraction. This method is in principle difficult with ground-based telescopes because of stability issues, but preliminary on-sky results with SPHERE show that the instrument stability is very high and is also allowing high contrast imaging  with reference stars. Contrasts of a few $10^{-6}$ can be feasible. The simulations here are assuming a favorable case, for which the atmospheric conditions are similar for both the target and the reference stars.

In the mid IR, we follow the procedure of \citet{bocc05}, to model MIRI coronagraphic images in several filters. MIRI offers phase mask coronagraphs at the shortest wavelengths, and a Lyot coronagraph (2" mask radius) at 23$\mu m$. The Four Quadrant Phase Mask \citep[FQPM, ][]{roua00} on MIRI allows to detect an object as close as the angular resolution. High contrast performance involves the use of a reference star observed in the same condition as the target star, an efficient method for space telescopes, which can be considered quite stable \citep{schn14}. The expected stability of JWST will allow principal component analysis with a sample of several reference stars to further improve the contrast. 

\subsubsection{Synthetic images}

As in Fig.~\ref{synthimage}, we consider the system in its ``asymmetric disc'' phase at $t \sim 10^{4}$ years.

With SPHERE, at 1.6$\mu m$ in scattered light, the contrast between the disc and stellar residuals is $\sim 10^{-5}$ in the peak luminosity region around the location of the initial release at 6\,AU, but it rapidly falls off to only $10^{-7}$ at 15\,AU. For a system viewed at 10pc, the simulated synthetic image is able to retrieve the inner part of the disc close to the inner ring at $\sim 6\,$AU, at about a factor 2 above the detection limit. The left/right asymmetry in this inner region is clearly visible, but the outer regions of the disc are undetectable (Fig.~\ref{SPHEREimage}a). The 30 pc distance case is less favourable: the disc is angularly resolved but fainter than the stellar residuals (Fig.~\ref{SPHEREimage}b). 

The situation is more favorable for MIRI in thermal emission (Fig.~\ref{MIRIimage}), because the disc/stellar contrast is enhanced at these longer wavelengths. At 30\,pc the disc is quite small compared to the angular resolution of MIRI ($>0.3''$). It is marginally detected at 11.4 and 15.5$\mu$m, but not at 23$\mu$m (the disc is too small compared to the Lyot mask size). By contrast, at 10\,pc the disc structures, in particular the location of the collision at 6\,AU, are well detected at all wavelengths.
The 11.4$\mu$m image at 10\,pc is relatively similar to the one obtained with the SPHERE simulation. It mainly shows the inner ring at the release distance (6\,AU) and clearly reveals the increased luminosity of the right-hand side (the release location) as compared to the left one (Fig.~\ref{MIRIimage}). The situation is basically the same at 15.5$\mu$m, except that the image is brighter, which is logical because we are here closer to the wavelength at which the disc luminosity peaks. The disc brightness is such that even a system 500 times fainter would still be above the detectability limit.
The peak luminosity is reached on the 23$\mu$m image, but the $2''$ Lyot coronagraph does here occult a large part of the disc. At 30\,pc, this corresponds to 60\,AU, meaning that the whole disc is hidden. At 10\,pc, the central shadowed region reduces to 20\,AU, so that the external regions of the post-breakup disc become visible. Interestingly, the brightest side of the disc is now the left region. This is a logical result, because this left outer region is mostly populated by small grains collisionally produced in the dense right-hand-side ansae of the ring and placed on eccentric orbits by radiation pressure (see Sect.~\ref{asymd}).

We conclude that multi-wavelength observations with SPHERE, and above all MIRI, have the potential to resolve the signature of massive collisional events in the inner regions of nearby debris discs.

\section{Parameter dependence, extrapolation to alternative set-ups} \label{param}

All simulations have been carried out for one specific set-up, and are thus strictly speaking only valid for this specific configuration. Because of the high-CPU cost of each LIDT-DD simulation, which is the price to pay for the full dynamics+collisions coupling, a thorough exploration of all the simulation's free parameters can unfortunately not be carried out. 
For some essential parameters, however, some test runs have been performed with alternative values. For all other parameters, we have to resort to simple scaling laws that allow to extrapolate, to a first order, our results to alternative set-ups. To derive these scaling laws, we mostly rely on the semi-empirical relations obtained by \citet{wyat07} and \citet{lohn08}, linking a debris disc's total luminosity to its main set-up parameters (radial location, total mass, dynamical state, etc...)

We stress, however, that such scaling laws should only be taken as rough indicators. Firstly because they were derived for axisymmetric discs and do not take into account the fact that, for example, the particle size distribution strongly varies with location (as a result of the coupling between collisions and radiation pressure), so that different scaling laws might apply depending on which region of the system is considered.
Another issue is that changing the set-up will often directly affect the fragment's initial velocity dispersion, which will in turn modify the physical outcome of collisions (rate of fragmentation, cratering and level of dust production) in a very non-linear way, as well as changing the shape of the inner ring containing most of the fragments' mass.

\subsection{Breakup related parameters}\label{bup}

\begin{figure}
\includegraphics[scale=0.36]{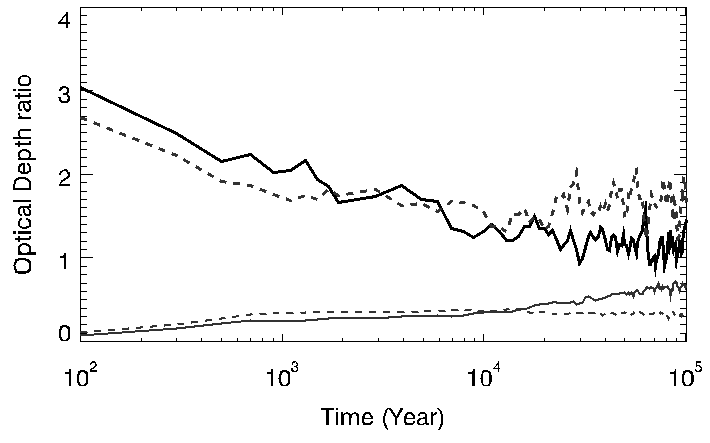}
\caption{Evolution of the system's left-right asymmetry, as quantified by the right/left Flux ratio within the inner ring (top) and in the region beyond it (bottom) for two different initial kick velocities: $v_\mathrm{esc}$ (nominal case, full line) and 2 $v_\mathrm{esc}$ (dotted line). }
\label{asym123}
\end{figure}

\begin{figure}
\includegraphics[scale=0.33]{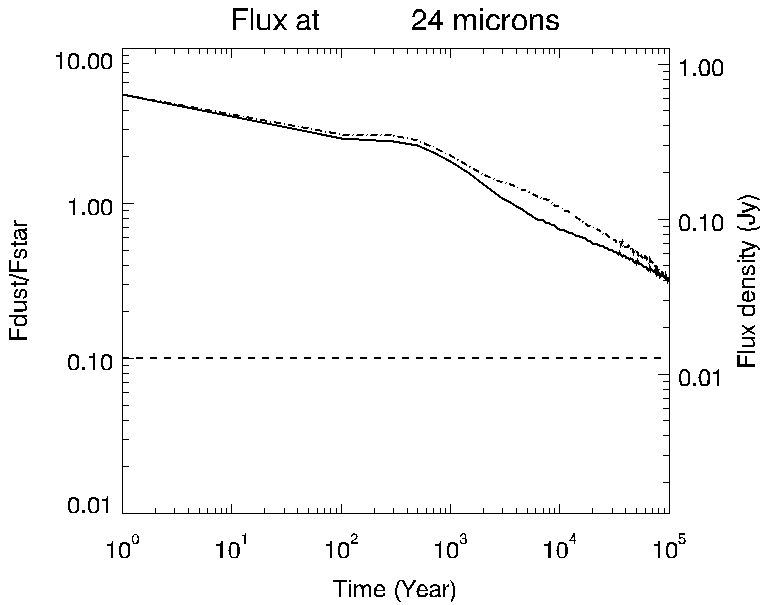}
\caption{Evolution of the dust's disc-integrated flux at 24 $\mu$m, as computed with GRaTer for two different initial kick velocities: full line is 1\,$v_\mathrm{esc}$ and dash-dotted 2\,$v_\mathrm{esc}$. The dashed line marks a dust-to-star flux ratio equal to 0.1, taken as our detectability criteria. The X-axis indicates the time after the breakup in years. The right-hand side Y-axis indicates the absolute flux in Jy, while the left-hand side axis displays the ratio of the disc flux to that of the stellar photosphere.}
\label{flux123}
\end{figure}

One crucial parameter that has been numerically explored is the initial velocity distribution of the post-breakup ejecta. As explained in Sect.~\ref{setup}, the $0\leq v_\text{frag}\leq v_\text{esc}$ choice was motivated by the fact that the most likely (or rather the least unlikely) target-shattering impacts are probably those having a kinetic energy just above the fragmenting threshold, and thus a low post-impact velocity dispersion.
However, other initial distribution of $v_\text{frag}$, even if less probable, cannot be ruled out. As a consequence, we have explored the system's evolution for a higher velocity dispersion, i.e., $0\leq v_\mathrm{frag}\leq 2\,v_\mathrm{esc}$. Figs.~\ref{asym123} and \ref{flux123} show the evolution of the left/right asymmetry and of the $24\mu$m flux for this alternative initial $v_\mathrm{frag}$ and for our nominal case. As can be seen, the asymmetry, after a similar initial resorption period, does then reach a plateau after a few $10^{3}$ years, after which its decrease is much slower than for the reference $v_\mathrm{frag}\leq v_\mathrm{esc}$ case. The reason for this behaviour is that higher values of $v_\mathrm{frag}$ lead to higher orbital eccentricities for the initially released fragments. For this higher $v_\mathrm{frag}$ dispersion, the ``inner ring'' formed by the largest fragments after a few 100 years, and which contains the bulk of the system's mass, becomes much wider on the left-hand side (and more diluted) than on the right-hand side (where all initial fragments have to pass through the initial point of release). Thus, the collisional production of small, high-$\beta$ grains is lower on the left-hand side than on the right-hand side of the ring, meaning that there are comparatively less small grains in the \emph{outer regions} of the right-hand side (since these are the grains produced in the left-hand side region of the ring). As a consequence, the asymmetric character of the system is maintained for the whole duration ($10^{5}$ years) of the run. 
Note, however, that the decrease of the dust-induced excess \emph{luminosity} is as fast as in the $v_\mathrm{frag}\leq v_\mathrm{esc}$ case (Fig.~\ref{flux123}) and follows the same evolution in $\sim t^{-0.3}$. So that the duration of the phase during which the system is detectable is expected to be approximately the same, that is, $t_\text{detect}\sim 10^6$ years.

Another important parameter is the maximum size $s_\text{max}$ considered for the initial ejecta' size distribution. As underlined by \citet{jack14}, this parameter is poorly constrained and could easily vary by several orders of magnitude depending on the initial impact's configuration. At any rate, our choice of an $s_\text{max}=1\,$m is probably an underestimation of the real value of $s_\text{max}$, which should probably be in the km-size range for the violent breakup of a Ceres-size body \citep{lein12}. However, increasing the value of $s_\text{max}$ from 1\,m to 1\,km (while keeping constant the total mass of ejecta $M_\text{frag}$) should not drastically alter our results. Indeed, because of the very steep size distribution in $dN \propto s^{-3.8} ds$ for the initial fragments, the quantity of $\leq\,1\,$mm dust released for a $s_\text{max}=1\,$km distribution is less than a factor $4$ smaller than for our nominal $s_\text{max}=1\,$m case. This would lead to a slightly lower dust-induced luminosity excess, but this would be (at least partially) compensated by a slower collisional evolution.

The last breakup related parameter is the total mass of ejecta. If we assume that, for the early ``spiral and ripples'' phase, the system's evolution is purely dynamical, then varying $M_\text{frag}$ will simply change the total luminosity during this stage following $L_\text{disc} \propto M_\text{frag}$. When the system enters its ``asymmetric disc'' phase and becomes collision-dominated, we then scale its temporal evolution by the variation of $t_\text{col}$. In principle, $t_\text{col}$ should scale as $M_\text{frag}^{-1}$, as can be seen in Equ.~36 of \citet{lohn08}. However, the variation of $M_\text{frag}$ also implies a change in $v_\text{frag}$, because we assume that $v_\text{frag}$ is close to the escape velocity of the shattered body of mass $M_\text{frag}$. Hence, $v_\text{frag} \propto M_\text{frag}^{1/3}$, and thus, according to Loehne's Equ.~36 once again, $t_\text{col} \propto M_\text{frag}^{-5/9}$.

\subsection{Disc-Star configuration}

The parameter that is the easiest to extrapolate is the distance $d$ at which the system is observed. Here, the results can be directly taken as they are, the only scaling being that of the total fluxes, which scale as $1/d^{2}$. The apparent sizes of the spatial structures as observed from Earth scale as $1/d$ (this has been explored to some extent in Sect.~\ref{images}).

Changing the location $r$ of the breakup will not directly affect $v_\text{frag}$ since the fragments' velocity dispersion is, in our assumption, only linked to the escape velocity of the shattered planetesimal. It will, however, affect the dynamical timescales, which scale as $r^{1.5}$, and the Keplerian velocities, which scale as $r^{-0.5}$. 
The $r^{1.5}$ scaling of the dynamical timescales can be used to directly infer the duration of the short-lived ``spiral and ripples'' phase (see Sect.~\ref{spiral}). 
The collisional timescale $t_\text{col}$ is $\propto r^{4.5}$ \citep[see Equ.~36 in][where we take $q_p \sim 2$ to account for our -3.8 primordial PSD slope]{lohn08}. This dependence is rather strong and could significantly change the duration of the asymmetric disc phase. As for the disc's luminosity, to a first-order, it scales as $L_\text{disc} \propto r^{-2}$ \citep{wyat08}, or if more accuracy is needed, new accurate fluxes can be quickly derived with a radiative transfer code such as GRaTer.

The change in $v_\text{Kep}$ will firstly affect the eccentricity distribution of the fragments (which is $\propto v_\text{frag}/v_\text{Kep}$) and thus the width of the inner ring at radial distance $r$. As a consequence, for the \emph{same shattering event}, we expect post-breakup discs to be more asymmetric at larger radial distances\footnote{Note, however, that this trend is, to a certain extent, balanced by the fact that planetesimal-shattering impacts should be, on average, less violent at larger distances because of the decreased $v_\text{Kep}$.}. The change in the initial eccentricity distribution will also affect $t_\text{col}$, which is $\propto e^{-5/3}$ \citep{lohn08}. Since $e \propto r^{0.5}$, we get $t_\text{col} \propto r^{-5/6}$. The $v_\text{Kep}\propto r^{-0.5}$ trend will also affect the outcome of physical collisions, in particular the fraction of shattered, cratered or accreted material. This effect is partially taken into account in Equ.~\ref{equlohn}, in the form of the smallest impactor that can fully shatter a given target, but many important consequences, in particular the level of cratered or re-accreted material cannot be easily extrapolated. 

Another important parameter is the stellar mass. $M_\star$ will moderately affect the dynamical timescales and collision rates, which both scale as $M_\star^{-0.5}$. Its biggest influence, however, is on the magnitude of the radiation pressure force, and in particular the value of the blow-out size $s_\text{cut}$, which is $\propto M_\star/L_\star$. Given that, in the $\sim 0.4-20M_{\odot}$ range, $L_\star$ is $\sim M_\star^{x}$, with $x$ in the $3.5-4$ range \citep{sala06}, we get that $s_\text{cut} \propto M_\star^{1-x}$. These strong variations of $s_\text{cut}$ as a function of $M_\star$ have important consequences. The main one is that they affect the size of the grains that dominate the system's cross section (which is always relatively close to $s_\text{cut}$), and thus the wavelengths at which the system's spatial features can be best observed. 
For low-mass stars, additional consequences are expected. There is in particular a critical stellar mass, typically in the $0.5-0.9M_\odot$ range, below which there is no longer a blow-out size since $\beta <0.5$ for all grains \citep[e.g.][]{reid11}. For these cases, the aftermath of the breakup will unfold in a very different way, and cannot be extrapolated from our reference simulation for an A star. Even solar-mass central stars should lead to a qualitatively different evolution, as only a small fraction of grains have $\beta >0.5$, i.e., only those close to the peak $\beta$ value, and many small grains below $s_\text{cut}$ become bound again because of the bell-shape of the $\beta(s)$ curve \citep[see for example Fig.~3 of][]{erte11}. 
As a consequence, we remain careful and conclude that the present results can only be extrapolated to cases with early-type stars for which there is large fraction of $s \leq s_\text{cut}$ grains on unbound orbits. Lastly, changing the stellar mass also implies changing the dust heating source, i.e.  the dust temperature and dust luminosity, and so the observational appearance of the dust cloud.

\subsection{Illustrative example: a pluto mass body colliding at 30AU}

\begin{figure*}
\makebox[\textwidth]{
\includegraphics[scale=0.25]{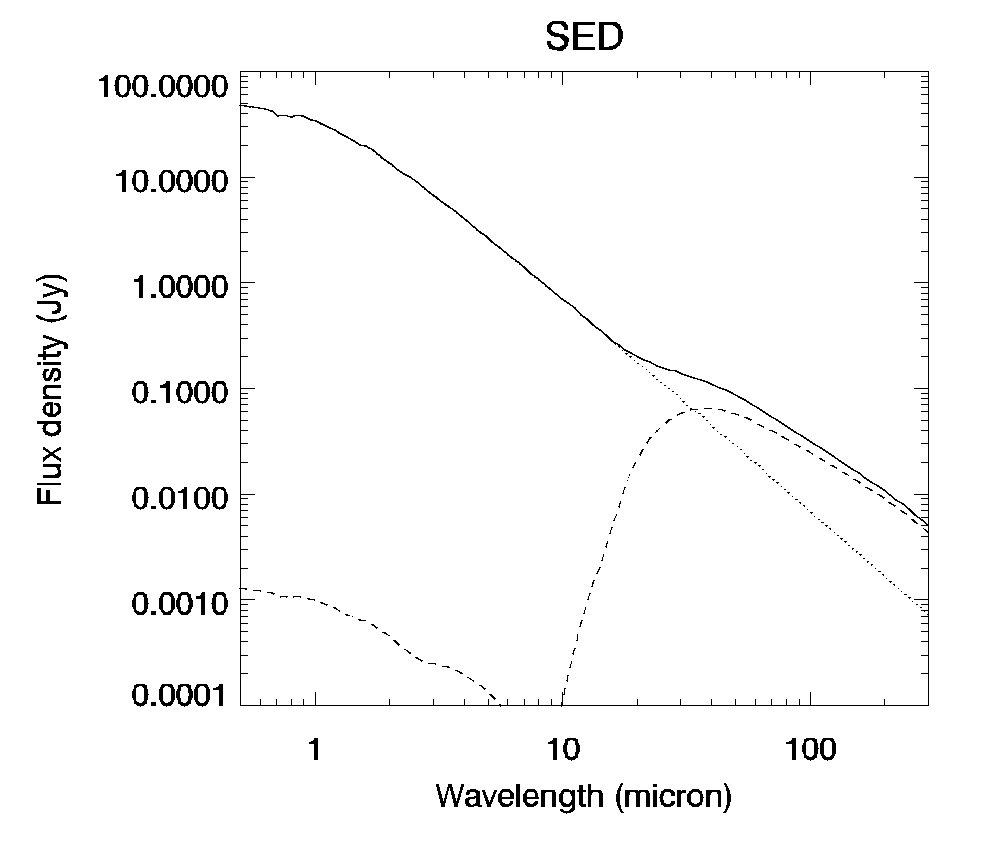}
\includegraphics[scale=0.34]{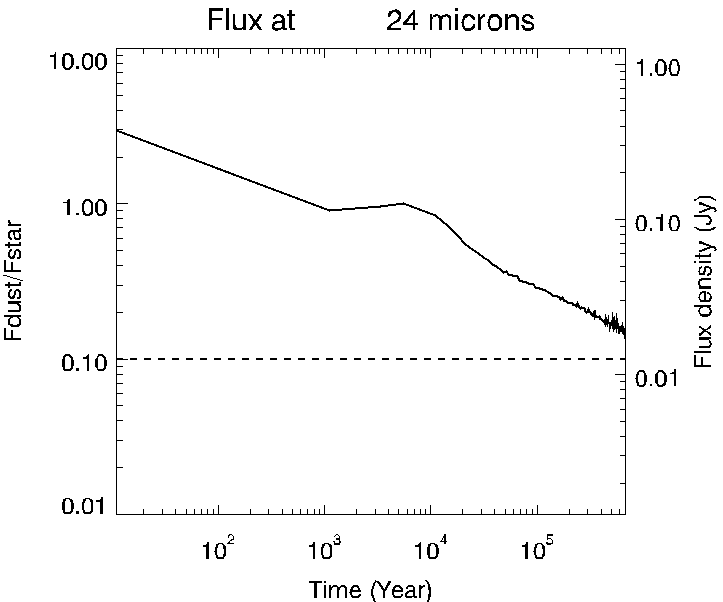}}
\caption{Case of a Pluto-mass body shattered at 30 AU around an A7V star, as observed from 30\,pc. Left: SED of the integrated system, as estimated with the GRaTer package at $7 \times 10^4$ years after the breakup. The full line represents the total (stellar photosphere + disc) luminosity, while the dashed line represents the sole contribution of the dust disc and the dotted one is that of the stellar photosphere. Right: Evolution of the disc-integrated flux at 24 $\mu$m, as computed with GRaTer (solid line). The dashed line marks a dust to star flux ratio equal to 0.1 taken as our detectability criteria.}
\label{pluto}
\end{figure*}

We illustrate these simple scaling laws with a specific example, the case of a Pluto-mass ($M_\text{pluto} \sim 1.3\times 10^{22}$ kg) body breaking up at $r_\text{new} =$ 30 AU (the stellar type remaining the same). 
To account for the change in mass, we follow the linear trend presented in Sect.~\ref{bup} and multiply the fluxes by the ratio $M_\text{pluto}/M_\text{frag} \sim 13$. Likewise, to account for the change in radial distance, all the particle positions have been rescaled by $r_\text{new}/r_\text{init} = 5$. As for the timescales, we rescale them by the ratio of the dynamical timescales, i.e., $(r_\text{new}/r_\text{init})^{1.5}\sim 11$, for the initial ``spiral and ripples'' phase, and by the ratio of the collisional timescales for the subsequent asymmetric-disc phase. Since $t_\text{col} \propto r^{4.5} M_\text{frag}^{-1} r^{-5/6} M_\text{frag}^{-5/9}= r^{11/3} M_\text{frag}^{-14/9}$, the collisional timescales are here increased by a factor $(r_\text{new}/r_\text{init})^{11/3}\times (M_\text{Pluto}/M_\text{frag})^{-14/9} \sim 6.8$. The GRaTer package is then used to work out the new temperatures of the grains (according to their new positions) and derive the new fluxes at each wavelength.

We plot in Fig.\ref{pluto}a the SED of the system, derived with GRaTer, during its asymmetric disc phase at $t \sim 7 \times 10^{4}$ years (to be compared with the nominal case at $10^{4}$ years showed on Fig.~\ref{SED}b). The release point being further out, the grain temperatures are lower and the SED accordingly peaks here around $40\mu$m.
Note that the flux density at 24$\mu$m is still high ($\sim$ 30 mJy), only three times smaller than for our reference case at 10kyrs. This is because of the larger mass of the shattered object, which partially compensates for the greater radial distance. We plot in Fig.~\ref{pluto}b the flux at 24$\mu$m versus time. As in our reference case, this flux stays above Spitzer/MIPS detectability limit for the whole duration of the simulation. Extrapolating the curve evolution, we find that this limit should be reached at $t \sim 3 \times 10^{6}$ years. As for the duration of the asymmetric phase, if we make the first order approximation that it is $\propto t_\text{col}$, we get that it is $\sim 7\times 10^5$ years (since $t_\text{col}$ is $\sim 6.8$ longer here).

\section{Limitations}\label{limit}

Apart from the high-CPU cost of the simulations, which prevent a thorough exploration of the parameter space, a potential limitation of the presented study is that it ignores other mechanisms that could affect the evolution of post-breakup discs. Among those are the presence of already formed giant planets, whose secular perturbations, and the differential precession they induce on orbiting objects, could change the rate at which the post-breakup asymmetries fade out.
For an exterior planet of mass $M_p$ having a semi-major axis $a_p$, the secular precession timescale is, in the limit of small $a/a_p$ \citep{murr99,must09}:
\begin{equation}
t_\text{sec}=\frac{4}{3} \frac{M_\star}{M_p} \left(\frac{a_p}{a}\right)^3 t_\text{orb},
\label{equsec}
\end{equation}
where $t_\text{orb}=2\pi\sqrt{a^3/GM_\star}$ is the orbital timescale for a particle of semi-major axis $a$.
For $t_\text{orb}=11$ years (which corresponds to our inner ring at 6 AU), a Jupiter mass planet at 30 AU would for instance randomize the longitude of the periastra on a timescale $t_\text{sec} \sim 3.5 \times 10^6$ years. This secular timescale is a factor 10 longer than the collisional one and should thus not drastically alter our nominal results. However, for alternative initial configurations, or if the giant planet is a lot closer, these secular effects could become non negligible. Note that, if needed by the astrophysical context (a specific disc+planet system for instance), LIDT-DD is designed to handle such pertubers.

Another limitation, related to the LIDT-DD code itself, is the finite resolution of the grid that is superimposed on the system in order to estimate its collisional evolution. For obvious CPU time constraints, the 2-D cells within which collisional interactions between all present super-particles are handled cannot be infinitely small. For the set-up considered here, this means that the grid is probably too coarse close to the location of the initial breakup, and that we probably underestimate the collisional rates in this narrow region (because collision rates are averaged over a cell that is too big). Thus the left/right contrast in terms of collision rates in the inner ring is probably underestimated. However, a test run with a refined grid having twice the resolution of the nominal one showed results which did not depart more than 10\% from the nominal case in terms of left/right asymmetries beyond $t\sim 10^{3}$years. This is probably because, for the limited fragment velocity dispersion considered in our nominal case, the singularity at the breakup location does not reach extreme levels \citep[as was for instance the case for the high $v_\text{frag}/v_\text{Kep}$ case considered by][]{jack14}. Another reason is that any initially extreme collisional singularity should be self-resorbing after a given time because of its highly eroding evolution.
We remain, however, conservative in our conclusions and consider that our runs give a lower limit for the duration of the asymmetric disc phase of the system's evolution.

\section{Conclusion and Perspectives}\label{discussion}

We have investigated the aftermath of the breakup of a large asteroid-like object in the inner regions of a planetary system, using for the first time a fully self-consistent model coupling the dynamical and collisional evolution of the system.
We have considered the case of the violent breakup of a single massive body in an otherwise dust-empty region. This configuration had already been investigated in some previous studies, but so far only with simplified collisional and/or dynamical prescriptions \citep{liss09,jack12,john12,jack14}. LIDT-DD allows us to relax most of these restrictive assumptions by self-consistently following the collisional and dynamical fate of the breakup fragments.

Our simulations have shown that the breakup of a Ceres-sized body at 6 AU from an A star leads to a luminosity excess that greatly exceeds that of a standard disc at collisional steady-state. The breakup's aftermath can be decomposed into three distinct phases.
At first, a bright spiral, composed of close-to $\beta=0.5$ grains, quickly forms and evolves into ripple-like structures over a few dynamical timescales. In parallel, a narrow ring, made of the largest breakup fragments, forms by Keplerian shear at the radial location of the release. The second phase, which is more long-lived, corresponds to an asymmetric disc, brighter and more compact on the side of the initial breakup and more extended and diffuse on the opposite side. The luminosity of this disc decreases with time as $\sim t^{-0.3}$, while its asymmetries are progressively resorbed by collisional activity. The third and final phase corresponds to the symmetrization of the system, which occurs on a timescales of a few $10^5$ years.
An important point is that asymmetries are here resorbed by collisional activity \textit{alone}, in the absence of planets or any other perturbing bodies or processes. More specifically, they are resorbed by the gradual dispersion of material due to the coupled effect of successive collisions and Keplerian motion, and the reprocessing of new fragments in regions initially devoid of material. 

Using the GRaTer package, we find that the flux excess created by the initial breakup should be clearly detectable, at $24\,\mu$m, for the reference case of a Ceres mass body at 6 AU from an A7V star. This luminosity excess should be observable in photometry, at 30\,pc, for at least $\sim 10^6$ years with Spitzer/MIPS (at 24$\mu$m).

To assess the observability of the asymmetries, we compute synthetic images for the SPHERE/VLT and MIRI/JWST instruments. With SPHERE at 1.6$\mu$m, the left/right asymmetry at the collision point (6 AU) should be detectable from a 10\,pc distance, but just above the detection limit. The situation is more favourable with MIRI, as the same asymmetry should be clearly seen, at 10\,pc, well above the detection limit at 11.4 and 15.5$\mu$m, and even be marginally visible at a distance of 30\,pc. At 23$\mu$m, because of the large Lyot coronagraph, only the external regions of the disc can be mapped out, but the left/right asymmetry is here also detected, albeit with the anti-breakup side now being the brightest. This is an expected result and could be used as an indicator of the signature of massive collisional events.

Due to the high CPU cost of each individual run, no full exploration of all the parameter space could be carried out in the present study. We derive, however, simple scaling laws that allow to extrapolate our results to alternative set-ups. Such simple laws should, however, only be regarded as rough first-order estimates and cannot reproduce all the results of LIDT-DD simulations.
The thorough LIDT-DD investigations of individual astrophysical cases exceeds the scope of the present work. Such dedicated studies, in particular that of the archetypal HD172555 system, will be the purpose of a forthcoming paper. The present study has, however, shown the potential of the LIDT-DD code for such future investigations, in particular when coupled to the GRaTer package to produce accurate SEDs and synthetic images.  

\begin{acknowledgements}

The authors would like to thank Mark Wyatt for pointing out to the limitation inherent to the finite size of the collisional grid, and Zoe Leinhardt for fruitful discussions on collisional prescriptions. The authors would like to thank the referee for fruitful comments that helped improve the manuscript. Q.K. acknowledges financial support from the French National Research Agency (ANR) through contract ANR-2010 BLAN-0505-01 (EXOZODI). SC acknowledges the financial support of the UnivEarthS Labex program at Sorbonne Paris CitŽ (ANR-10-LABX-0023 and ANR-11-IDEX-0005-02).

\end{acknowledgements}

{}

\clearpage

\end{document}